\documentclass[12pt,oneside,letterpaper]{article}
\usepackage{titlesec}
\titleformat*{\section}{\large\bfseries}
\titleformat*{\subsection}{\normalsize\bfseries}
\usepackage{amssymb}
\usepackage{amsmath}
\usepackage[dvips]{graphicx}
\usepackage{setspace}
\usepackage{stackrel}
\usepackage{amsfonts}
\usepackage{fancyhdr}
\usepackage{caption, subcaption}
\usepackage[dvipsnames]{xcolor}
\usepackage{caption}
\usepackage{graphicx}
\usepackage{rotating}
\usepackage{comment}
\usepackage{color}
\usepackage{cite}
\usepackage{braket}
\expandafter\def\csname ver@subfig.sty\endcsname{}

\definecolor{darkgreen}{rgb}{0,0.5,0}
\definecolor{darkblue}{rgb}{0,0,0.6}
\definecolor{purple}{rgb}{0.4,.2,0.7}
\newcommand{\p}{\partial}

\newcommand{\f}{\frac}

\newcommand{\be}{\begin{equation}}
\newcommand{\ee}{\end{equation}}

\usepackage{tikz}
\usetikzlibrary{matrix}
\usetikzlibrary{decorations.markings,calc,shapes,decorations.pathmorphing,patterns,decorations.pathreplacing}
\usetikzlibrary{positioning}

\definecolor{penroseblue}{RGB}{94,153,220}
\definecolor{penrosered}{RGB}{158,28,74}

\usepackage[colorlinks=true,citecolor=darkgreen,linkcolor=black,urlcolor=purple]{hyperref}

\usepackage{pdfsync}

\makeatletter
\newcommand*{\defeq}{\mathrel{\rlap{%
                     \raisebox{0.3ex}{$\m@th\cdot$}}%
                     \raisebox{-0.3ex}{$\m@th\cdot$}}%
                     =} 
\makeatother

\DeclareMathOperator{\Tr}{Tr}
\def\be{\begin{eqnarray}}
\def\ee{\end{eqnarray}}

\newcommand{\bea}{\begin{eqnarray}}
\newcommand{\eea}{\end{eqnarray}}

\def\ben{\begin{equation}}
\def\een{\end{equation}}

\let\a=\alpha \let\b=\beta \let\g=\gamma \let\d=\delta 
   
\let\l=\lambda \let\m=\mu \let\n=\nu  \let\r=v
\let\s=\sigma \let\t=\tau \def\ben{\begin{equation}}
\def\een{\end{equation}}

\let\a=\alpha \let\b=\beta \let\g=\gamma \let\d=\delta 
   
\let\l=\lambda \let\m=\mu \let\n=\nu  \let\r=v
\let\s=\sigma \let\t=\tau

  \let\D=\Delta  \let\L=\Lambda

\def\nn{\nonumber}
\let\f=\frac

\def\be{\begin{equation}}
\def\ee{\end{equation}}
\def\ba{\begin{array}}
\def\ea{\end{array}}

  \let\D=\Delta  \let\L=\Lambda

\def\nn{\nonumber}
\let\f=\frac

\def\be{\begin{equation}}
\def\ee{\end{equation}}
\def\ba{\begin{array}}
\def\ea{\end{array}}

\newcommand\edgar[1]{\textcolor{purple}{{\it #1 -Edgar}}}
\numberwithin{equation}{section}

\def\ba#1\ea{\begin{align}#1\end{align}}
\def\bs#1\es{\begin{split}#1\end{split}}

\usepackage{amsmath}

\renewcommand{\p}{\partial}

\usepackage[top=1in, bottom = 1in, left = 1in, right = 1in]{geometry}

\allowdisplaybreaks  

\begin{document}
\onehalfspacing

\begin{center}

~
\vskip5mm

{\LARGE  {
Timelike Liouville theory and \\ \vspace{2mm} AdS$_3$ gravity at finite cutoff 
}}

\vskip8mm

Kuroush Allameh and Edgar Shaghoulian

\vskip8mm
{ \it UC Santa Cruz\\
Physics Department\\
1156 High Street\\
Santa Cruz, CA 95064}


\end{center}

\vspace{4mm}

\begin{abstract}
\noindent
We propose that AdS$_3$ gravity with conformal boundary conditions is described by coupling the holographic CFT$_2$ to timelike Liouville theory and deforming by an exactly marginal operator. In this description, the Liouville field controls the finite-cutoff radial wall in the bulk. We check this proposal in the semiclassical limit by matching the sphere and torus partition functions  between the bulk and boundary theories. We also show that the Liouville field's equation of motion gives the bulk Hamiltonian constraint. The strong coupling limit of our theory pushes the bulk description deep inside the interior of black hole geometries. This is also the flat-space limit, and it leads to a duality between 3d flat space and 2d CFT.
\end{abstract}



\thispagestyle{empty}
\pagebreak
\pagestyle{plain}

\setcounter{tocdepth}{3}
{}
\vfill
\clearpage
\setcounter{page}{1}

\tableofcontents
\vfill\pagebreak

\section{Introduction}
One of the most surprising recent outputs of gravitational thermodynamics is the scaling  $S \sim T^{d-1}$ of the Bekenstein-Hawking entropy  of the $(d+1)$-dimensional Schwarzschild black hole \cite{Anninos:2023epi}. The key to this novel scaling is to use conformal boundary conditions in gravity, which fix the conformal class of the metric and the trace of the extrinsic curvature $K$ at the boundary \cite{York:1972sj, York:1986it, York:1986lje, Anderson:2006lqb, Witten:2018lgb}. (See also \cite{Bredberg:2011xw, Anninos:2011zn} for analysis of these boundary conditions in the fluid-gravity literature.) This generically leads to the cutoff living somewhere in the bulk of the spacetime. The scaling with temperature is remarkable because it is extensive in the boundary dimensionality.\footnote{We should dispel a common confusion, presumably stemming from a misreading of \cite{Susskind:1994vu, Susskind:1998dq}. The fact that the area-law of black hole entropy looks like a ``volume" law on the boundary is not sufficient to obtain extensivity of the entropy upon pushing the cutoff boundary close to the horizon. This is because our notion of extensivity requires the appropriate factors of temperature to appear as well, $S \sim V_{d-1} T^{d-1}$ for volume $V_{d-1}$. In fact, if you push the cutoff toward a fixed horizon in a high-temperature limit, then there is no way to get the high-temperature divergence $S \sim T^{d-1}$ needed for extensivity, since $S = A/(4G)$ remains finite. It is necessary instead that the horizon grows without bound in the high-temperature limit, as occurs with conformal boundary conditions \cite{Banihashemi:2025qqi}.} This is the same as the scaling in a local quantum field theory, and provides a tantalizing hint that there might be a $d$-dimensional local QFT description of the UV completion of the gravitational system, as in AdS/CFT. If true, then the tool of the thermal effective action of the boundary system makes several predictions for the thermodynamics, just as in AdS/CFT where they can be checked \cite{Allameh:2024qqp, Benjamin:2023qsc}. These predictions -- covering subextensive corrections to the entropy and coupling to different spatial backgrounds -- were corroborated for conformal boundary conditions in \cite{Banihashemi:2024yye}. Since the seminal work \cite{Anninos:2023epi}, this extensivity has been extended to de Sitter \cite{Anninos:2024wpy} and anti-de Sitter spacetimes \cite{Anninos:2024xhc, Banihashemi:2025qqi}, with the predictions of the thermal effective action confirmed and extended to inclusion of $U(1)$ charge and angular velocity \cite{Banihashemi:2025qqi}. 

When we have black hole solutions with both inner and outer horizons, conformal boundary conditions allow us to push the cutoff inside the black hole by scaling $K \rightarrow \infty$. This occurs by passing through a near-horizon near-extremal limit,  where the solution is governed by an extremal geometry with an AdS$_2$ factor \cite{Banihashemi:2025qqi}. This may allow a route to explore the geometry deep inside a black hole. (The usual instability of inner horizons does not seem to be relevant to this boundary value problem, and the solutions inside the inner horizon are necessary for thermodynamics consistent with the thermal effective action.)

A natural conjecture about these boundary conditions, explored in \cite{Banihashemi:2024yye, Banihashemi:2025qqi} to help explain the negativity of a certain Wilson coefficient of the thermal effective action \cite{Allameh:2024qqp}, is that the dual theory is a QFT coupled to a metric degree of freedom, the Weyl mode. This conjecture is simplest to explore in 3d gravity, where the Weyl mode is (almost) all there is to gravity in the 2d boundary. Thus in this paper we will analyze AdS$_3$ gravity with conformal boundary conditions. Fixing $K\ell > 2$ leads to geometries which have boundaries inside the usual asymptotic AdS boundary. The necessary boundary term for this variational problem leads to a Euclidean action
\be
I= -\f{1}{16\pi G} \int d^3x \sqrt{\mathcal{G}} \,(R+2/\ell^2) - \f{1}{16\pi G} \int d^2 x \sqrt{g} \,K\,.\label{cbcact1}
\ee
We use $\mathcal{G}_{\mu\nu}$ for the bulk metric to preserve $g_{\mu\nu}$ for the boundary metric. Notice the factor of $1/2$ relative to the usual Gibbons-Hawking-York boundary term needed for Dirichlet boundary conditions.  We will not add any boundary terms beyond the one written above. Interestingly, this action has no power-law divergences in the AdS/CFT limit $K\ell \rightarrow 2$ \cite{Banados:1998ys}. We can calculate a ``conformal" Brown-York stress tensor   \cite{York:1986lje, An:2021fcq}, obtained by varying the action \eqref{cbcact1} with respect to the conformal metric on the boundary $\overline{g}_{\m\n} = g^{-1/2} g_{\m\n}$, which is unique in a conformal class. This gives
\be\label{cbcstress1}
T^{CBC}_{\m\n} = -\f{2}{\sqrt{\hspace{.4mm}\overline{g}}} \f{\d I}{\d \hspace{.1mm}\overline{g}^{\,\m\n}} = \f{1}{8\pi G}(K_{\m\n} - \f 1 2 K g_{\m\n})\,.
\ee
The superscript CBC stands for conformal boundary conditions, reminding us that we use the action \eqref{cbcact1} to obtain it.  This stress tensor has a manifestly vanishing trace. The conformal metric satisfies $\sqrt{\hspace{.4mm}\overline{g}} = 1$ but we included this factor for uniformity with the ordinary Brown-York stress tensor.

We will concretely realize the idea in \cite{Banihashemi:2024yye, Banihashemi:2025qqi} by proposing a dual description in terms of coupling a two-dimensional CFT to timelike Liouville theory and deforming by an exactly marginal operator. 
We will show that this theory captures several aspects of the bulk physics. 

\subsection*{Summary of results}

We will begin by calculating the  sphere and torus partition functions in AdS$_3$ with conformal boundary conditions in Section \ref{spherebulk} and \ref{torusbulk}. The sphere partition function is found to be 
\be
\log Z[S^2] = \f{\ell}{4G} \log \f{K\ell + 2}{K\ell -2}\,.
\ee
This is independent of the size of the sphere and is therefore consistent with the vanishing of the trace of the stress tensor \eqref{cbcstress1}. The torus partition function has a richer structure. It has black hole solutions, which are isometric to a patch $r \in [r_+,r_c]$ of the rotating BTZ solution, and cosmic horizon solutions, which are isometric to a patch $r \in [r_c, r_-]$. The torus partition function takes exactly the required form of a CFT$_2$ \cite{Cardy:1986ie}, with a modular-invariant structure and an effective central charge \cite{Anninos:2023epi, Banihashemi:2024yye} 
\be\label{ceff}
c_{\text{eff}} = \f{3\ell}{2G}\,\f{K\ell - \sqrt{K^2\ell^2-4}}{2}\,.
\ee
We see that there is an explicit function of $K\ell$ which governs the modification of the Brown-Henneaux central charge $3\ell/2G$ \cite{Brown:1986nw}.   Together with the vanishing stress tensor trace $T^\m_\m = 0$, this indicates that the theory has a vanishing anomaly central charge and a nonvanishing effective central charge controlling the growth of the density of states. A vanishing anomaly central charge indicates that all stress tensor correlators vanish, which we will  confirm by a bulk calculation in Section \ref{correlators}. In Section \ref{transfbulk} we will revisit and modify the prescient analysis of \cite{Hartle:1983ai, Coleman:2020jte} to show how the path integral with conformal boundary conditions can be obtained by an integral transform of the path integral with Dirichlet boundary conditions.

The bulk analysis outlined above will lead us to a proposal for the holographic dual to this gravitational system. The dual is described by (a) coupling timelike Liouville theory (with central charge $c_L$) to the CFT$_2$ dual to AdS$_3$ gravity with asymptotic Dirichlet boundary conditions (with central charge $c_m = -c_L$) and (b) deforming by an exactly marginal operator, given by $\tilde T\tilde{\overline{T}}e^{-2\Phi}$ in the gravitational sector and in the semiclassical limit $c_m \rightarrow \infty$. The exact marginality of this operator is discussed in Appendix \ref{app:classmarg}. It is the gravitationally dressed $\tilde T\tilde{\overline{T}} =  \f 1 8(\tilde T_{\m\n}\tilde T^{\m\n}-(\tilde T_\m^\m)^2)$ operator \cite{Smirnov:2016lqw, Cavaglia:2016oda, McGough:2016lol}. The tildes represent quantities on the fiducial metric $\tilde{g}_{\m\n}$, defined in relation to the physical metric in the semiclassical limit as $g_{\m\n} = e^{2\Phi} \tilde{g}_{\m\n}$.  Our action for timelike Liouville theory in this limit is given by
\be\label{tlintro}
S_{tL} =\f{1}{4\pi \mathfrak{b}^2} \int d^2 x \sqrt{\tilde{g}} \left[-(\tilde{\nabla}\Phi)^2 - \tilde{R} \Phi + 4\pi \mathfrak{b}^2\m  e^{2\Phi}\right]
\ee
with $\mathfrak{b}^2 \approx 6/c_m$, while our deformation is 
\be\label{deformintro}
\f{\p S}{\p \l} = \int d^2 x \sqrt{\tilde{g}}\,\tilde T\tilde{\overline{T}}  e^{-2\Phi} + \f{c_m}{48\pi \l} \int d^2 x \sqrt{\tilde{g}} \tilde{R}\,.
\ee
This is the prescription for the gravitational sector in the bulk; additional bulk fields will require dressing  a more general operator along the lines of \cite{Hartman:2018tkw}. The boundary parameters are given in terms of bulk parameters as 
\be
\m = \f{K\ell - 2}{16\pi G \ell}\,,\qquad \l = 16\pi G \ell\,,\qquad c_m = -c_L = \f{3\ell}{2G}\,.
\ee
The stress tensor appearing in \eqref{deformintro} is of the matter sector plus the Wess-Zumino sector (by the Wess-Zumino sector we mean the first two terms of \eqref{tlintro}, which have their origin in the matter theory's anomaly) plus the deformation, i.e. of the full action excluding the cosmological constant term $\m e^{2\Phi}$. This corresponds to the Brown-York stress tensor in the bulk computed with Dirichlet boundary conditions, $T_{\m\n}^{DBC}$. The full stress tensor of the boundary theory corresponds to the conformal Brown-York stress tensor in the bulk, $T_{\m\n}^{CBC}$. 

In Section \ref{hconst}, we will show that the equation of motion for the Liouville field in the semiclassical limit is equivalent to the Hamiltonian constraint (for radial slicing) in the bulk. In Sections \ref{spherebdry} and \ref{torusbdry}, we will calculate the sphere and torus partition functions in the semiclassical limit of this theory. The results will give an exact match with the bulk expressions quoted above.

In the discussion section we will expand on various features of our proposed duality. We will consider the flat-space limit, obtained by scaling the Liouville parameters as $\m \l \rightarrow \infty$ while leaving the central charge untouched. This reproduces the sphere and torus partition functions in 3d Einstein gravity with vanishing cosmological constant. Interestingly, the dual description is now in terms of a modular-invariant CFT$_2$. We will also discuss this same limit of parameters as a way to push the cutoff toward the black hole singularity. Patches which try to include the causal singularity $r=0$ correspond to $\m < 0$, indicating an instability. Fixed-area quantities, a traditional observable in Liouville theories, will be shown to be closely related to observables in standard $T\overline{T}$-deformed theories \cite{Zamolodchikov:2004ce, Smirnov:2016lqw, Cavaglia:2016oda}. Finally, we will briefly discuss prospects for higher dimensions, de Sitter spacetimes, and extending our boundary theory to include the $bc$ ghosts of the non-critical string.

\section{Bulk analysis}

\subsection{Sphere partition function}\label{spherebulk}
We will begin our bulk analysis by fixing our boundary conformal class of metrics to have $S^2$ as a representative.  The solution that dominates the path integral with this boundary condition is assumed to be
\be
ds^2 =\f{dr^2}{1+r^2/\ell^2} + r^2 d\Omega^2\,.
\ee
At constant $r_c$ we have
\be\label{keqnsphere}
K\ell = 2 \sqrt{1 + \ell^2/r_c^2} \implies r_c = \frac{2\ell}{\sqrt{K^2\ell^2-4}}\,.
\ee
Notice also that we have a continuous family of solutions related by background $PSL(2,C)$ transformations. These all give the same  geometry, since they are isometries of $\mathbb{H}^3$.  A $PSU(2)$ subgroup corresponds to the rotations of the sphere that do not change the solution, whereas the other three generators change the solution. This spontaneous breaking is like what occurs in the standard treatment of nearly AdS$_2$ spacetimes \cite{Maldacena:2016upp}. This symmetry structure is mirrored in the Liouville description by the $PSL(2,C)$ invariance of timelike Liouville theory, where constant-dilaton saddles preserve the three $PSU(2)$ symmetries and spontaneously break the rest. One must therefore quotient by this $PSL(2,C)$ to obtain a finite answer for the sphere partition function. We will comment in the discussion about a local way of eliminating this divergence. 

The on-shell action is given by 
\be
\log Z = \f{1}{16\pi G} \int d^3 x \sqrt{\mathcal{G}} (R + 2/\ell^2) + \f{1}{16\pi G} \int d^2 x \sqrt{g} \, K\,.
\ee
Using $R = -6/\ell^2$ and integrating gives
\be\label{logZ1}
-\f{1}{4\pi G\ell^2} \int d^3 x \sqrt{\mathcal{G}} + \f{K}{16\pi G} \int d^2 x \sqrt{g} =\frac{1}{2G} \left(\ell \sinh ^{-1}\f{r_c}{\ell}-\f{r_c^2}{\ell} \sqrt{1+\ell^2/r_c^2}+\f{K r_c^2}{2}\right)\,.
\ee
Using \eqref{keqnsphere} we see that the latter two terms cancel. The modified coefficient for the boundary term with conformal boundary conditions was crucial for this cancellation. The final answer is
\be
\log Z[S^2] = \f{\ell}{4G} \, \log \f{K\ell+2}{K\ell-2}= \f{c_m}{6} \, \log \f{K\ell+2}{K\ell-2}\,,\label{sphereZ}
\ee
where we used the Brown-Henneaux central charge $c_m = 3\ell/2G$ \cite{Brown:1986nw}. The subscript $m$ is because we will eventually refer to this as the matter central charge. As discussed in \cite{Banihashemi:2024yye}, while the nonvanishing sphere partition function suggests a nonvanishing conformal anomaly, it is actually independent of the size of the sphere. The parameters that enter the logarithm are simply parameters that define the theory. This means that $(T^\m_\m)^{CBC} = 0$, indicating a vanishing conformal anomaly. This is consistent with the vanishing trace of the conformal Brown-York stress tensor \eqref{cbcstress1}. Explicitly calculating the stress tensor for this solution, we find $T^{CBC}_{\m\n} = 0$. 

Now we turn to possible complex solutions of our boundary value problem. One way to anticipate them is to notice the $\sinh^{-1}\f{r_c}{\ell}$ in \eqref{logZ1}, which indicates that a principal value was chosen. This occurred due to the bulk integral of $\sqrt{\mathcal{G}} = r^2/\sqrt{1+r^2/\ell^2}$, which needs branch cuts for complex $r$. Let us instead work in a more convenient coordinate system:
\be
ds^2 = d\rho^2 + \ell^2\sinh^2\f{\rho}{\ell}\,\,d\Omega^2\,,\qquad K \ell= 2 \coth \f{\rho_c}{\ell}\,.
\ee
Now, instead of the real contour $\rho \in [0, \rho_c]$, we can choose a complex contour that connects $\rho = 0$ to $\rho = \rho_c + i \pi n\ell$. This leaves the conformal class of boundary metrics and $K$ invariant. The partition function for one of these contours  is given by
\be\label{zcomplex}
 Z[S^2] = \exp\left[\f{\rho_c}{2G} + \f{ i \pi n\ell}{2G}\right] =  \exp\left[\f{c_m}{6} \, \log \f{K\ell+2}{K\ell-2}+ \f{i \pi n c_m}{3}\right]\,.
\ee
The on-shell actions of the complex saddles are consistent with taking $\rho_c \rightarrow \rho_c + i \pi n\ell$ in the on-shell action of the real saddle. This is the answer one gets by picking a simple complex contour that goes along the imaginary axis from $\rho = 0$ to $\rho = i \pi n\ell$, and then traverses in an entirely real direction from $\rho = i \pi n\ell$ to $\rho = \rho_c + i \pi n\ell$. Of course by holomorphicity this on-shell action is independent of the contour connecting these endpoints. 
\subsection{Torus partition function}\label{torusbulk}
In this section we will consider the conformal canonical ensemble at finite temperature and angular potential. The boundary conformal class of metrics therefore has a twisted torus as a representative. The relevant partition function  written in terms of a Hilbert space trace is 
\be \label{Z-conformal}
Z(\tilde{\b}, \tilde{\Omega}) = \Tr \exp\left[-\tilde{\b}(\tilde{H} - \tilde{\Omega} J )\right],
\ee
where we have defined $\tilde{H}$ as the generator of time translations in a coordinate $\tilde{\t}$ defined below. The potentials $\tilde\b$ and $\tilde{\Omega}$ are defined in terms of our boundary metric\footnote{There are three notions of temperature in this problem. The first is the ordinary Hawking (inverse) temperature, $\b$, which is the (inverse) temperature measured out at infinity. The second is the proper  (inverse) temperature at $r=r_c$, $\b_p = \b\sqrt{\mathcal{G}_{\t\t}} = \b \tilde{\L}$. The third is the proper conformal (inverse) temperature at $r=r_c$, $\tilde{\b} = \b/R = \b_p/\sqrt{\mathcal{G}_{\phi\phi}}$. In AdS spacetimes with $r_c = \infty$ we have $\b = \tilde{\b}$ due to the equal redshift in time and space components.}
\be\label{rotbdrymetric}
ds^2|_{\partial \mathcal{M}} = \tilde{\Lambda}^2(d\t^2 + R^2 d\phi^2 ) = \Lambda^2(d\tilde{\t}^2 + d\phi^2),
\ee\vspace{-7mm}
\be\label{rotbdrymetric2}
 (\t,\phi)\sim(\t+\b, \phi + i \b \Omega),\quad (\tilde{\t},\phi) \sim (\tilde{\t} + \tilde{\b}, \phi + i \tilde{\b}\tilde{\Omega}),\quad \tilde{\b} = \b/R,\quad \tilde{\Omega} = R \Omega.
\ee
We also have the relation $\tilde{H}  \equiv \Lambda H^{\rm CBC}$, where $H^{\text{CBC}}$ is the boundary Hamiltonian for the action appropriate for conformal boundary conditions

There are two solutions which contribute to this partition function. The first is a patch of the spinning BTZ geometry \cite{Martinez:1999qi}, whose geometry is given by:
\be\label{rotbtzmet}
ds^2 = -f(r) dt^2 + \f{ dr^2}{f(r)}+r^2 \left(d\theta - \f{r_-r_+}{\ell r^2}dt\right)^2\,,\qquad f(r) = \f{(r^2-r_+^2)(r^2-r_-^2)}{\ell^2 r^2} 
\ee
\be
(t, \theta) \sim (t + i\b, \theta + i \b \Omega) \sim (t, \theta + 2\pi)\,.
\ee
The other solution is a patch of the thermal AdS geometry, given by setting $r_-=0$ and $r_+ = i \ell$ in the above geometry.

\begin{figure}[t]
  \centering \vspace{-10mm}
\begin{subfigure}{0.45\textwidth}
        \centering 
{\begin{tikzpicture}[scale=1.5]
\node[rotate=90,white] at (0,2.925) {\LARGE$\cdots$};
\node[rotate=90,white] at (0,-2.925) {\LARGE$\cdots$};

\draw[draw=none,pattern = crosshatch,pattern color=penrosered] (1,1) .. controls (0.45,0.25) and (0.45,-0.25) .. (1,-1) to (0,0) -- cycle;

\draw[-] (-1,-1) to (-1,1);
\draw[-] (1,-1) to (1,1);
\draw[-,penrosered!80!black,dashed,very thick] (-1,1) to (0,0) to (-1,-1);
\draw[-,penrosered!80!black,dashed,very thick] (1,1) to (0,0) to (1,-1);
\draw[-,decorate,decoration={zigzag,amplitude=0.5mm,segment length=2.5mm}] (-1,-1) to (1,-1);
\draw[-,decorate,decoration={zigzag,amplitude=0.5mm,segment length=2.5mm}] (-1,1) to (1,1);
\node[penrosered!80!black] at (0.35,0.7) {\large$r_+$};
\node[penrosered!80!black] at (0.35,-0.7) {\large$r_+$};

\draw[-] (1,1) .. controls (0.45,0.25) and (0.45,-0.25) .. (1,-1);

\node at (0.775,0) {\large$r_c$};

\node at (0,-1.5) {};
\node at (0,1.5) {};
\end{tikzpicture}}
        \caption{}
    \end{subfigure}
    \hfill
    \begin{subfigure}{0.45\textwidth}
        \centering
\begin{tikzpicture}[scale=1.35]
\draw[draw=none,pattern = crosshatch,pattern color=penroseblue] (1,3) .. controls (0.45,2.25) and (0.45,1.75) .. (1,1) to (0,2) -- cycle;
\draw[draw=none,pattern = crosshatch,pattern color=penrosered] (1,1) .. controls (0.45,0.25) and (0.45,-0.25) .. (1,-1) to (0,0) -- cycle;

\draw[-] (-1,-1) to (-1,1);
\draw[-] (1,-1) to (1,1);
\draw[-,penrosered!80!black,dashed,very thick] (-1,1) to (0,0) to (-1,-1);
\draw[-,penrosered!80!black,dashed,very thick] (1,1) to (0,0) to (1,-1);
\draw[-,penroseblue!80!black,dashed,very thick] (-1,1) to (1,3);
\draw[-,penroseblue!80!black,dashed,very thick] (1,1) to (-1,3);
\draw[-,penroseblue!80!black,dashed,very thick] (-1,-1) to (1,-3);
\draw[-,penroseblue!80!black,dashed,very thick] (1,-1) to (-1,-3);
\draw[-,decorate,decoration={snake,amplitude=0.5mm,segment length=3.8mm}] (-1,1) to (-1,3);
\draw[-,decorate,decoration={snake,amplitude=0.5mm,segment length=3.8mm}] (1,3) to (1,1);
\draw[-,decorate,decoration={snake,amplitude=0.5mm,segment length=3.8mm}] (-1,-3) to (-1,-1);
\draw[-,decorate,decoration={snake,amplitude=0.5mm,segment length=3.8mm}] (1,-1) to (1,-3);
\node[penrosered!80!black] at (0.35,0.7) {\large$r_+$};
\node[penrosered!80!black] at (0.35,-0.7) {\large$r_+$};

\node[penroseblue!80!black] at (0.35,2+0.7) {\large$r_-$};
\node[penroseblue!80!black] at (0.35,2-0.7) {\large$r_-$};

\node[rotate=90] at (0,3.25) {\LARGE$\cdots$};
\node[rotate=90] at (0,-3.25) {\LARGE$\cdots$};

\draw[-] (1,3) .. controls (0.45,2.25) and (0.45,1.75) .. (1,1);
\draw[-] (1,1) .. controls (0.45,0.25) and (0.45,-0.25) .. (1,-1);

\node at (0.775,2) {\large$r_c$};
\node at (0.775,0) {\large$r_c$};

\node at (-1.75,0) {};
\node at (1.75,0) {};
\end{tikzpicture}
 \caption{}
    \end{subfigure}

\caption{(a) The Penrose diagram for the non-rotating BTZ geometry. Only black-hole type solutions, represented by the red patch, are possible. (b) The Penrose diagram for the rotating BTZ geometry. We can construct both black-hole patches (in red) and  cosmic patches (in blue). The wavy curves at $r=0$  indicate the different nature of the singularity.}
\label{figs:penroseAdS}
\end{figure}
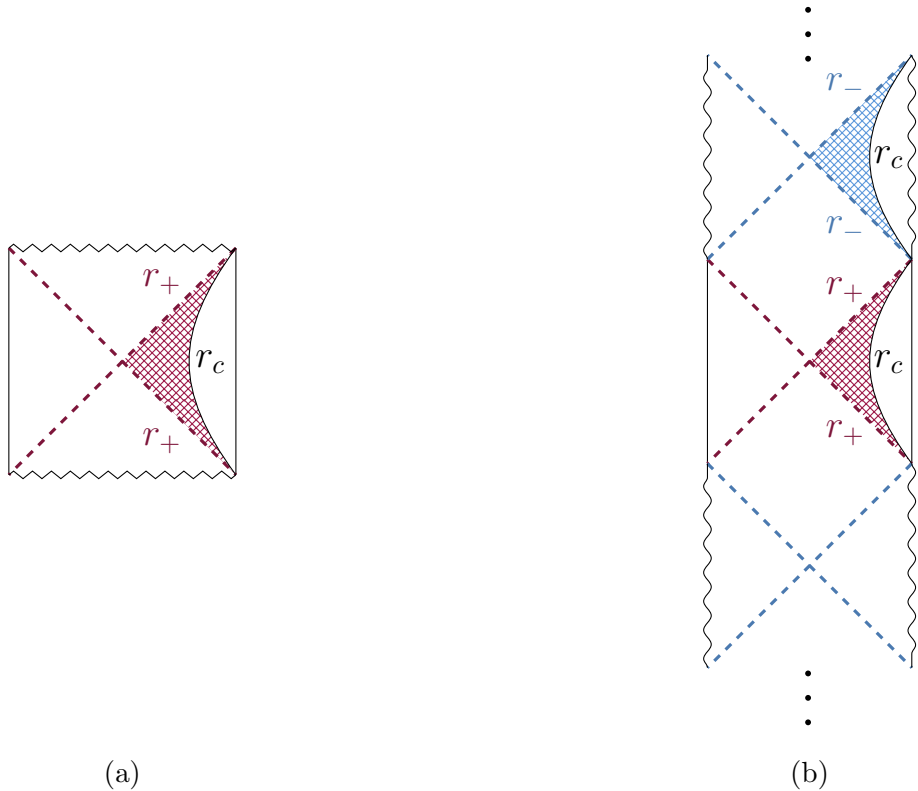

We will begin by analyzing the BTZ geometry. The non-rotating BTZ geometry has a single horizon, see Figure \ref{figs:penroseAdS}(a). The relevant solutions to our conformal canonical ensemble are patches  $r \in [r_+, r_c]$. The singularity at $r=0$ is a Milne-type singularity. The spinning BTZ geometry has an outer and an inner horizon, which means in our conformal canonical ensemble we can have Euclidean solutions corresponding to patches $r \in [r_+, r_c]$, which we call black hole solutions, and Euclidean solutions corresponding to patches $r \in [r_c, r_-]$, which we call cosmic solutions. See Figure \ref{figs:penroseAdS}(b). The singularity at $r=0$ marks the region beyond which we have closed timelike curves, since $t$ and $\phi$ exchange character for $ r^2 < 0$; generic couplings to matter will turn this into a genuine curvature singularity \cite{Banados:1992gq}. 

We should comment about the analytic continuation required to obtain Euclidean solutions. For the black hole patches $r \in [r_+, r_c]$, this is standard \cite{Gibbons:1976ue}. One takes $t \rightarrow i\t$, which pops out a factor of $i$ in the $d\theta dt$ cross-term, although the metric nevertheless has three positive eigenvalues. Sometimes this is followed by an analytic continuation in the angular velocity, although we will not do this. We will instead consider the complex metric as a solution to the boundary-value problem with real angular velocity (and by some abuse of language continue to refer to it as a Euclidean solution). The reason we do this is because the continuation required for the cosmic patch $r \in [r_c, r_-]$ is slightly different. First, one must shift $\theta \rightarrow \theta + r_+ t/(\ell r_-)$ before continuing $t \rightarrow i\t$ to ensure we get a metric with three positive eigenvalues. Recall that this shift is fixed by enforcing the cross-term to vanish at the (in this case, inner) horizon. In fact for the black hole patch we should really do this as well, but we get a metric with three positive eigenvalues regardless. Inside the inner horizon, the $t-\theta$ mixing is too strong. The second difference is that we cannot get rid of the $i$ in the metric with an analytic continuation in a chemical potential without changing the signature of the metric. As a result, to deal with both patches uniformly, we will shift the $\theta$ coordinate and then analytically continue in $t$, ending with a complex metric. The thermodynamic parameters resulting from this procedure are given by the standard formulas
\be
\b = \pm \f{4\pi}{f'(r_\pm)}\,,\quad \Omega = \f{r_\mp}{\ell r_\pm}\,,
\ee
where we used $\Omega = \f{\mathcal{G}_{t\theta}}{\mathcal{G}_{\theta\theta}}|_{r_\pm}$. The upper signs are for black hole solutions, where we impose regularity at the outer horizon, whereas the lower signs are for cosmic solutions, where we impose regularity at the inner horizon. To write down the appropriate boundary values of the thermodynamic parameters, we pick a cutoff $r_c$ and define a new angular coordinate $\phi = \theta - \f{r_-r_+}{\ell r_c^2} t$, which gives an induced metric
\be
ds^2 = - f(r_c) dt^2 + r_c^2 d\phi^2\,,\qquad (t, \phi) \sim \left(t+ i \b,\phi + i \b \left(\Omega - \f{r_- r_+}{\ell r_c^2}\right)\right).
\ee
Notice that the shifted angular velocity $\Omega - r_- r_+/(\ell r_c^2)$ is simply subtracting off the rotation at the cutoff $r_c$. This term is usually absent in AdS/CFT with $d=2$ since the angular velocity at $r_c = \infty$ vanishes. To correctly identify the boundary potentials in the conformal boundary condition problem, we need to put the metric in the form \eqref{rotbdrymetric}-\eqref{rotbdrymetric2}. To do so, we simply identify $\tilde{\Lambda} = \sqrt{f(r_c)}$, $R = r_c/\sqrt{f(r_c)}$ (which also means $\Lambda = r_c$) to get 
\be
\tilde{\b} = \b/R = \b \f{\sqrt{f(r_c)}}{r_c}\,,\qquad \tilde{\Omega} = \f{r_c}{\sqrt{f(r_c)}} \left(\Omega - \f{r_- r_+}{\ell r_c^2}\right)\,.
\ee
The relevant equations are therefore
\be\label{rotbtztriple}
\tilde{\beta} = \pm\frac{4\pi}{f'(r_h)} \frac{\sqrt{f(r_c)}}{r_c},\quad K = \pm\frac{1}{r_c \sqrt{f(r_c)}}\left[f(r_c) + \frac{r_c f'(r_c)}{2}\right],\quad \tilde{\Omega} =\pm\f{r_\mp r_c }{ r_\pm\ell \sqrt{f(r_c)}} \left(1-\f{r_\pm^2}{r_c^2}\right),
\ee
where the $\pm$ in the equation for $\tilde{\Omega}$ was inserted to restrict to $\tilde{\Omega} \geq 0$. This freedom is due to the $\tilde{\Omega} \rightarrow -\tilde{\Omega}$ symmetry, coming from flipping the sign of $\mathcal{G}_{t\theta}$ in \eqref{rotbtzmet}. As before, the upper signs correspond to black hole solutions while the lower signs correspond to cosmic solutions.

These boundary parameters have interesting bounds. We can write
\be
\tilde{\Omega}^2 = \f{ r_{\mp}^2 - \f{r_+^2 r_-^2}{r_c^2}}{r_\pm^2-\f{r_+^2 r_-^2}{r_c^2}} \leq 1\,,
\ee
where the inequality follows from the fact that in the black hole case $r_- < r_+ < r_c$, while for the cosmic horizon case $r_c < r_- < r_+$. 

The equation for $K$ can be written as 
\be
K \ell = \pm \f{(r_c-r_+)(r_c+r_+)+(r_c-r_-)(r_c+r_-)}{\sqrt{(r_c-r_+)(r_c+r_+)(r_c-r_-)(r_c+r_-)}}  \geq 2\,,
\ee
which is simply the AM-GM inequality. Notice that this is true \emph{both} for black hole and cosmic horizon solutions. This is unlike $d > 2$, where one could access $K\ell<d$, which is less than the value given at the asymptotic AdS boundary $K\ell = d$, by going inside the inner horizon \cite{Banihashemi:2022jys}. Altogether we have
\be
\tilde{\b} > 0\,,\qquad K\ell \geq 2\,,\qquad 0 \leq \tilde{\Omega} \leq 1\,.
\ee
It is notable that we have the same bound on the (conformal) angular velocity $0 \leq \tilde{\Omega} \leq 1$ as occurs in the ordinary AdS/CFT problem. In both cases the upper bound is saturated for the extremal black hole $r_- = r_+$. This suggests the existence of a healthy dual, spinning subluminally.

We  now return to our equations \eqref{rotbtztriple}, which we solve to get 
\be\label{ranswers}
r_{\mp} = \f{2\pi \ell \tilde{\Omega}}{\tilde{\b}(1-\tilde{\Omega}^2)}\,,\quad r_\pm = \f{\pi \ell \left(K\ell - \sqrt{K^2\ell^2-4}\right)}{\tilde{\b}(1-\tilde{\Omega}^2)}\,,\quad r_c = \f{\pi\ell}{\tilde{\b}\sqrt{1-\tilde{\Omega}^2}} \sqrt{\f{2K\ell}{\sqrt{K^2\ell^2-4}}-2}\,.
\ee
 As before, the upper signs correspond to the black hole solution while the lower signs correspond to the cosmic horizon solution. Only one of these solutions exists for a given set of the parameters:
\begin{align}
K\ell &> \f{1+\tilde{\Omega}^2}{\tilde{\Omega}}  \implies \text{cosmic horizon solution}\,,\\
K\ell &< \f{1+\tilde{\Omega}^2}{\tilde{\Omega}} \implies \text{black hole solution}\,.
\end{align}
This also means that we have a near-horizon near-extremal limit as $K\ell \rightarrow (1 + \tilde{\Omega}^2)/\tilde{\Omega}$, where the cutoff surface moves toward a coalescing outer and inner horizon. This admits a dimensionally reduced JT gravity description \cite{Banihashemi:2025qqi}, see also \cite{Galante:2025tnt}.

Using our action \eqref{cbcact1} and solution \eqref{ranswers}, the on-shell action of the BTZ geometry is given by
\be\label{logzfull}
\text{BTZ:}\qquad \log Z(\tilde{\b},\tilde{\Omega}) = \f{\pi^2 \ell(K\ell - \sqrt{K^2 \ell^2-4})}{4G\tilde{\b}(1-\tilde{\Omega}^2)} =\f{\log Z(\tilde{\b})}{1-\tilde{\Omega}^2}\,,
\ee 
which implies an entropy
\be
S = (1-\tilde{\b}\p_{\tilde{\b}}) \log Z(\tilde{\b},\tilde{\Omega}) =   \f{\pi^2 \ell(K\ell - \sqrt{K^2 \ell^2-4})}{2G\tilde{\b}(1-\tilde{\Omega}^2)} =\f{2\log Z(\tilde{\b})}{1-\tilde{\Omega}^2}\,.
\ee
 This entropy formula is consistent with the area law $S = 2\pi r_\pm/4G$, upon using the relevant solution for $r_\pm$ from \eqref{ranswers}.

Now we turn to the thermal AdS geometry:
\be\label{thermalads}
ds^2 = (1+r^2/\ell^2)d\t^2 + \f{dr^2}{1+ r^2/\ell^2} + r^2 d\phi^2\,,\qquad (\t, \phi)\sim (\t + \tilde{\b}, \phi + i \tilde{\b}\tilde{\Omega})\,. 
\ee
As usual this geometry can be obtained from the family of BTZ metrics by setting $r_- = 0$ and $r_+ = i\ell$. The extrinsic curvature at the cutoff surface $r_c$ is given by
\be\label{kvac}
K = \f{1+2r_c^2/\ell^2}{r_c \sqrt{1+r_c^2/\ell^2}}\,.
\ee
Solving this for $r_c$ and using it in the evaluation of the action \eqref{cbcact1} gives
\be\label{zads}
\text{Thermal AdS:}\qquad \log Z(\tilde{\b}, \tilde{\Omega}) = \f{\tilde{\b}\ell}{8 G} \, \f{K \ell - \sqrt{K^2 \ell^2-4}}{2}\,.
\ee
Comparing to the BTZ on-shell action \eqref{logzfull}, we see that we have a transition at the modular-invariant locus $\tilde{\b}^2(1-\tilde{\Omega}^2) = 4\pi^2$:
\be
\log Z(\tilde{\b}, \tilde{\Omega}) =\begin{cases} \dfrac{\pi^2 \ell}{2G \tilde{\b} (1-\tilde{\Omega}^2) }\,\dfrac{K\ell - \sqrt{K^2 \ell^2-4}}{2} \,,\qquad \tilde{\b}^2(1-\tilde{\Omega}^2) < 4\pi^2 \\[1.5em]
\dfrac{\tilde{\b}\ell}{8 G} \, \dfrac{K \ell - \sqrt{K^2 \ell^2-4}}{2}\,,\qquad \qquad \qquad\tilde{\b}^2(1-\tilde{\Omega}^2) > 4\pi^2
\end{cases},
\ee
with the BTZ geometry dominating at high temperatures.

 We can calculate the thermal energy and  angular momentum in the high-temperature phase as 
\begin{align}\label{thermalans}
 \tilde{E}  &= -\p_{\tilde{\b}} \log Z(\tilde{\b},\tilde{\theta}) = \f{\pi^2\ell(1+\tilde{\Omega}^2) (K\ell - \sqrt{K^2\ell^2-4})}{4 G \tilde{\b}^2(1-\tilde{\Omega}^2)^2}\,,\\
 \tilde{J}  &= i\p_{\tilde{\theta}} \log Z(\tilde{\b},\tilde{\theta})  = \f{\pi^2 \ell \tilde{\Omega} (K \ell- \sqrt{K^2\ell^2-4})}{ 2G \tilde{\b}^2(1-\tilde{\Omega}^2)^2}\,,
\end{align}
where we used $\tilde{\theta} = i \tilde{\b}\tilde{\Omega}$. We see from these expressions that $ \tilde{E}  \geq  \tilde{J} $ with equality if and only if  $\,\tilde{\Omega} = 1$, i.e. the extremal black hole. As usual, $\tilde{E}$ is the expectation value of the Hamiltonian generating translations in $\tilde{\t}$, while $\tilde{J}$ is the expectation value of the momentum operator generating translations in $\phi$. The charges can be calculated from the conformal Brown-York stress tensor \eqref{cbcstress1} and give (using $J = \int dx\sqrt{h} \,u^\m T_{\phi\m}^{CBC}$ and $E = \int dx \sqrt{h} \, u^\m u^\n T_{\m\n}^{CBC}$)
\be
E = \pm\f{r_c^2(r_+^2+r_-^2)-2 r_+^2 r_-^2}{8 G \ell r_c \sqrt{(r_c^2-r_+^2)(r_c^2-r_-^2)}}\,,\qquad J = \f{r_+ r_-}{4G\ell}\,.
\ee
Using \eqref{ranswers} shows that $E = r_c^{-1} \tilde{E}$  and $J = \tilde{J}.$ The relation $E = r_c^{-1}\tilde{E}$ indicates that $E$ is conjugate to the proper temperature at finite cutoff, whereas $\tilde{E}$ is conjugate to the proper conformal temperature at finite cutoff, which is made dimensionless by a factor of $r_c$. The range of energies for these solutions is $r_c E \in (0,\infty)$ for both branches. An intriguing fact in the non-spinning case is that if we instead compute the quasilocal mass as $M = \int dx \sqrt{h} u^\m t^\n T^{CBC}_{\m\n}$ for timelike Killing vector $t^\n = u^\n \sqrt{f(r_c)}$, then we get an answer that is independent of the cutoff $r_c$. This mass corresponds to a choice of time $t$ in metric $-f(r_c)dt^2 + \dots$. This means that the combination $(K\ell - \sqrt{K^2\ell^2-4})/2$ simply equals $\sqrt{f(r_c)}/r_c$, which is the same as $\tilde{\b}/\b$. We will not use this in what follows, although a similar observation is true for black branes in arbitrary dimension.

\subsubsection*{Modular invariance}
As we mentioned above, the transition between vacuum and BTZ geometries happens along the fixed locus of the modular $S$ transformation. To see this, recall how the modular $S$ transformation behaves for a CFT$_2$ on a torus of spatial circle size $2\pi R$:
\be\label{Stransf}
Z(\b, \Omega) = Z\left(\f{4\pi^2R^2}{\b(1-R^2\Omega^2)}, 
\,\,-\Omega\right).
\ee
As $\b \rightarrow 0$ the dual channel indicates that we project to the ground state, which we assume has vanishing angular momentum and energy $E_{\text{vac}}=-c/12$:
\be\label{vacZ}
\log Z(\b, \Omega) \approx \f{\pi^2 c R}{3\b(1-R^2\Omega^2)} = \f{\pi^2 c }{3\tilde{\b}(1-\tilde\Omega^2)}\,.
\ee
In the final equation we used our definition $\tilde{\b} = \b/R$ and $\tilde\Omega = R \Omega$. Notice that our convention is that we divide the inverse temperature by the radius of the spatial circle (not its total length) to arrive at an inverse conformal temperature. The fixed locus of the modular $S$ transformation is when the arguments on both sides of \eqref{Stransf} are equal, which occurs at $\tilde{\b}^2 (1-\tilde{\Omega}^2) = 4\pi^2$.

By comparing \eqref{vacZ} with \eqref{logzfull}, we identify the effective central charge of our putative boundary theory as
\be\label{ceff}
c_{\text{eff}} = \f{3 \ell(K\ell - \sqrt{K^2 \ell^2 - 4})}{4G}\,.
\ee
We call this the effective central charge since as we have already seen the anomaly central charge vanishes.

The claimed modular invariant structure requires $E_{\text{vac}} = -c_{\text{eff}}/12$, so let's make sure this is true. We calculate the energy of global AdS$_3$ with conformal boundary conditions. The geometry is given by \eqref{thermalads}, and the dimensionless vacuum energy is calculated to be
\be
\tilde{E}_{\text{vac}} = -\f{r_c}{8 G  \sqrt{1+r_c^2/\ell^2}} = -\f{\ell(K\ell - \sqrt{K^2\ell^2-4})}{16 G}\,,
\ee
where in the final expression we used \eqref{kvac}. Equating this energy with $-c_{\text{eff}}/12$ again identifies the effective central charge as in \eqref{ceff}. Alternatively, we can equate the free energy \eqref{zads} with $ -\tilde{\b}\tilde{E}_{\text{vac}}$ and write $E_{\text{vac}} = -c_{\text{eff}}/12$ to extract the effective central charge, which also gives \eqref{ceff}.

As in ordinary AdS/CFT, we can get the energies on the cylinder $E \in (-c_{\text{eff}}/12, 0)$ by considering quotients of the global AdS$_3$ spacetime, i.e. conical deficits. This works simply because the quotient cuts out part of the space and therefore reduces the magnitude of the energy.  Negative-mass BTZ solutions are  conical deficit spacetimes for $-1 < M < 0$ and conical surpluses for $M < -1$. We exclude the latter as usual for giving energies below our ground state energy. In ordinary AdS/CFT, these surplus geometries would correspond to operators with negative scaling dimension. In our dual theory, we will see that all $M < 0$ states correspond to operators with negative scaling dimension, although the ones with $M\geq -1$ are admissible.

\subsection{Transforming between conformal and Dirichlet ensembles}\label{transfbulk}
In \cite{Hartle:1983ai}, a transform from Dirichlet boundary conditions (DBC) to conformal boundary conditions (CBC) was introduced in the context of the wavefunctional in quantum gravity. The basic idea is that $K$ and $\sqrt{g}$ are canonically conjugate variables, so in Lorentzian signature we have a usual Fourier transform to change bases from the position variables $\sqrt{g}$ to the momentum variable $K_L$, i.e. a factor of $e^{i \int d^d x \sqrt{g} K_L}$. The analytic continuation to Euclidean signature was considered, in which case $K_L = iK_E$ since the extrinsic curvature was being evaluated on a spacelike surface. Furthermore, the boundary condition $\Psi[\sqrt{g} < 0] = 0$ is imposed, leading to a semi-infinite contour which agrees precisely with the Laplace transform. As we will discuss below, $K_L$ cannot be self-adjoint so this formula should not be interpreted as a change of basis.  

In \cite{Coleman:2020jte}, the Laplace transform from \cite{Hartle:1983ai} was used to transform between Dirichlet and conformal boundary conditions for the partition function relevant to radial evolution in AdS$_3$. In this case, $K_L = K_E$ and the factor of $i$ comes instead from the measure $d^d x$ which has a factor corresponding to time $dt = i d\t$. 

An important caveat when interpreting the formulas below from a quantum-mechanical perspective is the following. Since $\sqrt{g}$ is semi-infinite, its conjugate $K$ cannot be a self-adjoint operator in the quantum theory. This is the same reason why there is no self-adjoint time operator in quantum theory, since the Hamiltonian is bounded below. Let's understand this in terms of an analogy to a particle on a semi-infinite line, $x \geq 0$. One simple way to see the non-adjointness of the conjugate $p$ is to consider a putative $p$-eigenstate and transform it to a position-space wavefunction $\psi_p(x) = \langle x|p\rangle$, which would satisfy
\be
-i \psi_p'(x) = p \psi_p(x) \implies \psi_p(x) = \f{1}{\sqrt{2\pi}} e^{i px}\,.
\ee
This putative wavefunction is a standard plane wave and therefore does not satisfy the boundary condition $\psi(x < 0) = 0$. What is worse, by calculating the deficiency index we can see that there is no self-adjoint extension of the $p$ operator.  In effect this means that the radial wavefunctions (which we call partition functions) are not to be thought of as an overlap $\langle K|\psi\rangle$, since there is no complete basis of $K$-eigenstates.\footnote{A canonical transformation one can consider in such a situation is $X = \log x$, $P = x p$, preserving the Poisson bracket $\langle X, P\rangle  = \langle x, p\rangle = 1$. This leads to variables $\log \sqrt{g}$ and $\sqrt{g} K$, and we can consider a boundary condition that fixes $\sqrt{g}K$. Recall the Fourier transform relating bases also indicates the appropriate boundary term in the classical variational problem that fixes $\sqrt{g}K$: $\int d^d x \sqrt{g} K \log \sqrt{g}$. This is the $p=1/2$ case of the family of boundary conditions considered in \cite{Liu:2024ymn}. This case has $S(\beta) \sim \b$, as indicated by black hole thermodynamics with these boundary conditions.} 

We can, however, write a transform between  thermodynamic ensembles instead of quantum-mechanical bases. In that case we can define an ``iso-$K$" or ``isocurvature" ensemble that keeps fixed $K$ and the conformal class of metrics. We can write down the transform based on the difference between the boundary terms for Dirichlet boundary conditions and conformal boundary conditions. But let's do it another way first to exhibit the consistency between the quantum-mechanical Fourier transforms and the boundary terms needed in the variational principle. To work out the conjugate to $\sqrt{g}$, we consider a variation of the volume mode of the metric by decomposing $\sqrt{g}\,\tilde{g}_{ij} =  g_{ij}$ and considering a variation with $\d \tilde{g}_{ij} = 0$:
\be
\d \sqrt{g} \,\tilde{g}_{ij} = \d g_{ij} \implies   \d g_{ij} = \f{g_{ij}}{\sqrt{g}} \d \sqrt{g}
\ee
We plug this into the variation
\be
\d S = \int d^2 x \,\pi^{ij} \d g_{ij}
\ee
where
\be
\pi^{ij} = \f{1}{16\pi G} \sqrt{g} (K^{ij} - g^{ij} K)
\ee
to get
\be
\pi^{ij} \d g_{ij} = -\f{K}{16\pi G} \d \sqrt{g}\,.
\ee
This identifies the conjugate to $\sqrt{g}$ as $-K/(16\pi G)$, which means that our transform in Lorentzian signature should be written as
\be
Z^L_{CBC}[\tilde{g}_{\m\n}, K] = \int D \sqrt{g} \, e^{i\, \f{1}{16\pi G} \int d^2 x \sqrt{g}K } \, Z^L_{DBC}[g_{\m\n}]\,.
\ee
In Euclidean signature $t \rightarrow i\t$ we get
\be
Z_{CBC}[\tilde{g}_{\m\n}, K] = \int D \sqrt{g} \, e^{-\, \f{1}{16\pi G} \int d^2 x \sqrt{g}K } \, Z_{DBC}[g_{\m\n}]\,.
\ee
Picking a spacetime-independent $\sqrt{g}$, this just becomes the formula in \cite{Hartle:1983ai, Coleman:2020jte}. For the AdS$_3$ case considered in \cite{Coleman:2020jte}, this is still not quite right. The partition function with Dirichlet boundary conditions is calculated with the inclusion of a cosmological constant counterterm. The partition function with conformal boundary conditions does not have such a term, as it would violate the boundary conditions (as mentioned before the modified coefficient of the Gibbons-Hawking-York term accomplishes the same regularization in the $K\ell \rightarrow 2$ limit as the cosmological constant counterterm). We therefore need to subtract this counterterm. Altogether this gives
\be\label{cbctransform2d}
Z_{CBC}[\tilde{g}_{\m\n}, K] = \int_{\tilde{g}_{\m\n}\,\,\text{fixed}}\hspace{-5mm} Dg\, e^{-I_{CBC}} = \int D\sqrt{g} \, Z_{DBC}[g_{\m\n}] e^{-\f{1}{16\pi G\ell }\int d^2 x\sqrt{g} \, (K\ell-2)}\,.
\ee
Notice that this result could have been obtained by considering the difference in actions between Dirichlet and conformal boundary conditions, as alluded to earlier: 
\be
I_{DBC} = I_{EH} - \f{1}{8\pi G} \int d^2 x \sqrt{g}\,(K-1/\ell)\,,\qquad I_{CBC} = I_{EH} - \f{1}{16\pi G  }\int d^2x \sqrt{g}\, K\,.
\ee\vspace{-5mm}
\be\label{actiondiff}
\implies I_{CBC} - I_{DBC} =  \f{1}{16\pi G\ell}\int d^2x \sqrt{g}\, (K\ell-2)\,.
\ee
The definition \eqref{cbctransform2d} agrees with the on-shell evaluation of the path integrals, due to \eqref{actiondiff}. For spacetime-independent $K$ and $\sqrt{g}$, the transform can more simply be written as 
    \be\label{mastertransform}
    Z_{CBC}[\tilde{g}_{\m\n}, K] = \int dV Z_{DBC}[V]e^{-\f{K\ell-2}{16\pi G\ell} V}\,.
    \ee
The integral is over the overall volume $V = \int d^2 x \sqrt{g}$ of the metric.

To see that this is consistent with transforms between ensembles using ordinary thermodynamics, let's consider a finite-temperature ensemble with spacetime-independent $K$ in the 2d boundary theory. We can calculate 
\be
Z(\tilde{\b}, K) =\int dV'\, \int d\tilde{U}  \rho(\tilde{U},V') e^{-\tilde{\b}(\tilde{U} + K V')}\,.
\ee
$\tilde{U}$ is what appears in the first law $d\tilde{U} = \tilde{T} dS - K dV'$, and $V'=V/(16\pi G \tilde\b)$ with $V$ equal to the spacetime volume (see Section 4 of \cite{Banihashemi:2024yye}; $V'$ was called $\tilde{V}$ in \cite{Banihashemi:2024yye}, but this would clash with our notation here). We define the partition function at fixed $\tilde\b$ and fixed volume $V'$ as
\be
Z(\tilde{\b}, V') = \int d\tilde{U}\, \rho(\tilde{U},V') e^{-\tilde{\b} \tilde{U}},\label{dbcz}
\ee
exactly in analogy to e.g. a fixed-charge thermodynamic ensemble. $Z(\tilde{\b}, V')$ is like the bulk partition function with Dirichlet boundary conditions, where the full metric is fixed. Plugging into the above gives 
\be\label{cbcz}
Z(\tilde{\b}, K) = \int dV'\, Z(\tilde{\b}, V') e^{-\tilde{\b} K V'}\,.
\ee
Up to the cosmological constant counterterm, this is consistent with \eqref{mastertransform} upon using $V'=V/(16\pi G \tilde\b)$. 

\section{Boundary analysis: deformed timelike Liouville theory}
Let us highlight some facts about the putative boundary theory as calculated from the bulk in the previous section:
\begin{itemize}
  \item The sphere partition function is given by 
    \be
\log Z[S^2] = \f{c_m}{6} \log \f{K\ell+2}{K\ell-2}\,,\qquad c_m = \f{3\ell}{2G}\,,
    \ee
    where $c_m$ is the usual Brown-Henneaux central charge. This formula, or the conformal Brown-York stress tensor \eqref{cbcstress1}, implies a vanishing stress tensor trace
    \be
    (T^\m_\m)^{CBC} = 0\,.
    \ee
    \item The theory on a torus exhibits modular invariance and a Cardy density of states with an ``effective" central charge 
    \be
    c_{\text{eff}} = \f{3\ell}{2G} \,\f{K\ell - \sqrt{K^2\ell^2-4}}{2}\,.
    \ee
    \item The bulk finite cutoff theory with Dirichlet boundary conditions can be related to the one with conformal boundary conditions by integrating its partition function over the Weyl mode, with kernel 
    \be
    Z_{CBC}[\tilde{g}_{\m\n}, K] = \int D\sqrt{g} \, Z_{DBC}[g_{\m\n}] e^{-\f{1}{16\pi G\ell }\int d^2 x\sqrt{g} \, (K\ell-2)}\,.
    \ee
\end{itemize}
The vanishing stress tensor trace of the first bullet point can be accommodated by a CFT with vanishing conformal anomaly. The second bullet point then requires that the minimum-dimension operator has negative scaling dimension and therefore sits below the identity operator. These two facts together already suggest a role for timelike Liouville theory. This interpretation is bolstered by the third bullet point, which takes some massive theory $Z_{DBC}[g_{\m\n}]$ and integrates it (with a kernel) over the volume mode to produce the conformal boundary condition partition function. The kernel term is simply a classical (as opposed to quantum-generated) cosmological constant term, which we can identify with the cosmological constant term in the Liouville action  \cite{David:1988hj, Distler:1988jt}
. The other two terms of Liouville theory comprise the Wess-Zumino anomaly effective action, which comes from the variation of the conformal mode in the integral above. Altogether, these facts suggest the following proposal: 
\newpage
\begin{quote}
Quantum gravity in AdS$_3$ with conformal boundary conditions is holographically dual to a 2d conformal field theory obtained by coupling the matter CFT$_2$ of AdS$_3$/CFT$_2$ to timelike Liouville theory and deforming by an exactly marginal operator. This operator is given by the gravitationally dressed $\tilde T\tilde{\overline{T}}$ operator. 
\end{quote}
To flesh out this proposal, we should establish our conventions. Since we will be interested in the semiclassical limit $c_m \rightarrow \infty$, we are in the arena of timelike Liouville theory (for which one just needs $c_m >25$).\footnote{The relevance of timelike Liouville theory to quantum gravity began with \cite{Polchinski:1989fn}. The semiclassical limit emphasized there has been pursued in \cite{Anninos:2021ene, Muhlmann:2022duj, Anninos:2024iwf}, with aspects of the Lorentzian Hilbert space studied in \cite{Anninos:2024iwf}. A formula for the three-point function, known as the timelike DOZZ formula (named after its spacelike cousin \cite{Dorn:1994xn, Zamolodchikov:1995aa}), was proposed from a bootstrap perspective in \cite{Zamolodchikov:2005fy, Kostov:2005kk}, see also  \cite{Schomerus:2003vv}. This formula was analyzed from the path integral perspective \cite{Harlow:2011ny} and a Coulomb gas representation \cite{Giribet:2011zx}.  Work towards a rigorous definition of the theory includes \cite{Harlow:2011ny, Ribault:2015sxa, Bautista:2019jau, Bautista:2020obj, Kapec:2020xaj, Guillarmou:2023exh, Chatterjee:2025yzo, Chatterjee:2026zmb}. Applications of the theory to higher-dimensional models of quantum gravity include \cite{Freivogel:2006xu, Collier:2023cyw}, and applications to probability theory and statistical mechanics include \cite{Delfino:2010xm, Ikhlef:2008zz, Ang:2021tjp}.} To access the semiclassical limit, we should define our theory at finite $c_m$. We do so by first writing our physical metric as $g_{\m\n} = e^{2\mathfrak{b} \varphi} \tilde{g}_{\m\n}$. The metric $\tilde{g}_{\m\n}$ is called the fiducial metric. Our action is
\be\label{tLaction}
S_{tL} = \f{1}{4\pi} \int d^2 x \sqrt{\tilde{g}} \left[-(\tilde{\nabla}\varphi)^2 -q \tilde{R} \varphi + 4\pi \m e^{2\mathfrak{b}\varphi}\right].
\ee
The minus sign in front of the kinetic term is responsible for much of the difficulty in handling this theory. The Hilbert space consists of vertex operators $e^{2\a\varphi}$ with $\a \in \mathbb{R}$. These states are normalizable. The reality of $\a$ implies a bound on the dimensions of the vertex operators:
\be
\text{Normalizable states}:\quad \D[e^{2\a\varphi}] = 2 \a(q+\a) \geq -\f{q^2}{2} = \f{c_{L}-1}{12}\,,
\ee
where $c_L$ is the central charge of this theory, given in \eqref{paramsl}. One implication of the above is that operators of negative scaling dimension are normalizable and exist in the spectrum. This will actually be crucial for our correspondence. To ensure that the interaction term $\m e^{2\mathfrak{b}\varphi}$ is exactly marginal, we have 
\be\label{iddress}
\D[e^{2\mathfrak{b}\varphi}]= 2\mathfrak{b}(q+\mathfrak{b})=2\,\,\implies\,\, \mathfrak{b} = \f 1 2\left(-q + \sqrt{4+q^2}\right)\,.
\ee
Altogether, the parameters of this theory are given by
\be\label{paramsl}
q = - \mathfrak{b} + 1/\mathfrak{b}\,,\qquad c_{L} = 1- 6q^2 \,.
\ee
The wrong-sign kinetic term of timelike Liouville theory can be thought of as the residual boundary piece of the wrong-sign kinetic term in the bulk Euclidean gravity path integral \cite{Gibbons:1978ac}. 

Now we want to deform this theory. The coupling to Liouville theory ensures exact conformal invariance, so any matter operator we use must be dressed by the Liouville field to become exactly marginal \cite{Distler:1988jt, Seiberg:1990eb}. A simple example of this is in the undeformed Liouville theory \eqref{tLaction}, where the identity operator is dressed with $e^{2\mathfrak{b}\varphi}$. For a more general operator $\mathcal{O}$ with dimension $\D$ we dress as $\mathcal{O} e^{2\a\varphi}$ and demand marginality via
\be\label{tLmarginal}
\text{Marginal dressing:} \qquad \D + 2 \a (q+\a) = 2\implies \a = \f 1 2 \left(-q + \sqrt{4-2\D + q^2}\right).
\ee
We picked the positive branch for the square root for consistency with the dressing of the identity operator \eqref{iddress}. To ensure we end with a normalizable state $\a \in \mathbb{R}$, we require $q^2 \geq 2 \D-4$. We see that all relevant operators $\D < 2$ satisfy this automatically. The more irrelevant the operator, the larger $q$ we need to ensure the constraint is satisfied. This condition on $\a$ ensures that the operator $\mathcal{O} e^{2\a\varphi}$ is exactly marginal in the undeformed Liouville theory. However, if we deform the theory by this marginal operator to trigger a renormalization group flow, the marginality is generally spoiled beyond tree level. This is a fundamental difficulty in defining  such a theory beyond the semiclassical limit; we will discuss this in Appendix \ref{app:qtm}.

A dramatic simplification occurs in the semiclassical limit 
\be\label{tlsemi}
\mathfrak{b} \rightarrow 0: \qquad \varphi = \mathfrak{b} ^{-1} \Phi\,,\qquad q \approx \f{1}{ \mathfrak{b} }\,,\qquad c_m = -c_L \approx 6 q^2 \approx \f{6}{\mathfrak{b} ^2}\,.
\ee
In this limit our undeformed timelike Liouville theory becomes
\be
S_{tL} =\f{1}{4\pi \mathfrak{b}^2} \int d^2 x \sqrt{\tilde{g}} \left[-(\tilde{\nabla}\Phi)^2 - \tilde{R} \Phi + 4\pi \mathfrak{b}^2\m  e^{2\Phi}\right].\label{timelikesemi}
\ee
Following \cite{McGough:2016lol}, we want to deform by the (suitably dressed) operator $\mathcal{O} = \tilde T\tilde{\overline{T}} = \f 1 8 (\tilde T^{\m\n}\tilde T_{\m\n} - (\tilde T^\m_\m)^2)$. The tildes on the stress tensors indicate that we consider the operator on the fiducial metric. As discussed in the introduction, $\tilde{T}_{\m\n}$ is the stress tensor of the theory \emph{excluding} the cosmological constant term.  Our proposal is that the dual theory for given $K\ell$ lives on the fixed line defined by 
\be
\f{\p S}{\p \l} = \int d^2 x \sqrt{\tilde{g}}\,\tilde T\tilde{\overline{T}} e^{-2\Phi} + \f{c_m}{48\pi \l} \int d^2 x \sqrt{\tilde{g}} \tilde{R}\label{semideform}
\ee
and parameter identifications
\be\label{paramids}
\m = \f{K\ell-2}{16\pi G \ell}\,,\qquad \l = 16\pi G \ell\,,\qquad c_{L} = -c_m = -\f{3\ell}{2G}\,.
\ee
In the semiclassical limit and with homogeneous $\Phi$, dressing the irrelevant, flowed $T\overline{T}$ operator to make it marginal corresponds to taking $\l \rightarrow \l e^{-2\Phi}$ and replacing the physical metric with the fiducial metric.

At linear order in the deformation, our action is simply
\be\label{lineardef}
 S_m[\tilde{g}] + \f{c_m \chi}{12}\log \l + \f{1}{4\pi \mathfrak{b}^2} \int d^2 x \sqrt{\tilde{g}} \left[-(\tilde{\nabla}\Phi)^2 - \tilde{R} \Phi + 4\pi \mathfrak{b}^2\m  e^{2\Phi} + 4\pi \mathfrak{b}^2 \l \tilde T\tilde{\overline{T}} e^{-2\Phi}\right].
\ee
Analyzing the theory at first order in $\l$ is particularly straightforward since the matter and Liouville sectors are decoupled at zeroeth order in $\l$. 

It is important to note that the total theory has $c = 0$, consistent with the manifest vanishing of the trace of the stress tensor in an arbitrary background. This means that the full stress tensor of the theory is both a primary and a descendant, i.e. it is null. It is not BRST exact, as occurs on the string worldsheet, since we do not gauge the diffeomorphisms and therefore do not introduce the $bc$ ghosts. (We will briefly comment on the introduction of $bc$ ghosts in Section \ref{discsec}.) This means the stress tensor remains a physical operator in the spectrum. Its $n$-point functions with itself, however, must vanish -- we verify this with a bulk argument in Section \ref{correlators}. In a unitary theory, the Cauchy-Schwarz inequality would dictate that the state created by a null operator decouples from the physical spectrum, since $|\langle T|\chi\rangle|^2 \leq \langle T|T\rangle \langle \chi|\chi\rangle = 0$ and $\langle T|T\rangle =  0$ together imply  $\langle T|\chi\rangle = 0$ for arbitrary physical state $|\chi\rangle$. 



As an aside, we point out that the continuous spectrum of pure timelike Liouville theory can be rendered discrete by considering the identification \cite{Kapec:2020xaj,Guillarmou:2023exh} $\varphi \sim \varphi + 2\pi i  r$, under which the linear-in-$\varphi$ term gives an invariance in $e^{-S}$ as long as $q\chi r \in \mathbb{Z}$, for $\chi$ the Euler character of the manifold. Considering manifolds with and without boundary, we have that $\chi \leq 2$ can be any integer. Thus we have $qr \in \mathbb{Z}$. The exponential term remains invariant if $2 b r \in \mathbb{Z}$. These two constraints require $b^2$ to be rational. Writing $b^2 = p/n$ in lowest terms, i.e. gcd$(p,n)=1$, we find that the minimal possible $r$ is $\sqrt{pn}/(\text{gcd}(n-p,2p))$. In the semiclassical limit this becomes $\Phi \sim \Phi + \pi i k$ for $k \in \mathbb{Z}$ and $c_m k/12 \in \mathbb{Z}$, which  requires rational $c_m$. The minimal $k$ is given by $c_mk/12 = w$ for minimal positive integer $w$. We will not actually perform this identification. 

Defining the quantum theory at finite $c_m$ is much thornier, see Appendix \ref{app:qtm}.


\subsection{Liouville equation of motion and the Hamiltonian constraint}\label{hconst}

In this section we will show that the Liouville equation of motion in the semiclassical limit is identical to the (radial) Hamiltonian constraint in the bulk. To do so, let's first write the Hamiltonian constraint for radial slicing:
\be\label{hconsbulk}
K^2-K_{\mu\nu}K^{\mu\nu}-\f{2}{\ell^2}-R^{(2)}=0,
\ee
where $K_{\mu\nu}$ is the extrinsic curvature on the slice and $R^{(2)}$ is the Ricci scalar of the slice. We can rewrite the extrinsic curvature in terms of the Brown-York stress tensor for Dirichlet boundary conditions:
\be\label{tdbc}
T^{DBC}_{\mu\nu}=\f{1}{8\pi G}\left(K_{\mu\nu}-K g_{\mu\nu}+\frac{g_{\mu\nu}}{\ell}\right).
\ee
This lets us rewrite \eqref{hconsbulk} as
\be\label{traceeq}
(T^{\mu}_\mu)_{DBC}=-4\pi G\ell\left(T_{\mu\nu}^{DBC}T^{\mu\nu}_{DBC}-(T^{\mu}_\mu)_{DBC}^2\right)-\frac{\ell R^{(2)}}{16\pi G}.
\ee
This is sometimes called the trace flow equation in the context of the $T\overline{T}$ deformation. Now consider the $\Phi$ equation of motion for the semiclassical boundary theory defined by \eqref{timelikesemi}-\eqref{semideform}:
\be\label{eomhcons}
2\tilde{\square}\Phi-\tilde{R}+8\pi\mathfrak{b}^2\mu e^{2\Phi}-8\pi \mathfrak{b}^2\lambda \tilde T\tilde{\overline{T}}(\lambda)e^{-2\Phi}=0.
\ee
This equation of motion is derived in Appendix \ref{app:ttbarwz}. In the semiclassical limit we have $ds^2=e^{2\Phi}d\tilde{s}^2$, so we can write
\be\label{physfid}
R=e^{-2\Phi}(\tilde{R}-2\tilde{\square}\Phi),\quad T\overline{T}(\lambda)=e^{-4\Phi}\tilde T\tilde{\overline{T}}(\lambda)\,.
\ee
In this expression $T\overline{T}(\l)$ is constructed from the matter stress tensor on the physical metric, whereas $\tilde T\tilde{\overline{T}}$ is constructed from the matter plus Wess Zumino stress tensor on the fiducial metric; this means that the anomaly is included in going from physical to fiducial metric in \eqref{physfid}.\footnote{\label{footnote5}The equation $T\overline{T}(\l) = e^{-4\Phi} \tilde T \tilde{\overline{T}}(\l)$ can be written with the dependence on the volume manifest, $T\overline{T}(\l, V) = e^{-4\Phi} \tilde T \tilde{\overline{T}}(\l, \tilde V)$. (There can be additional dependencies we leave out, e.g. on the temperature.) $V$ is the physical volume and $\tilde{V}$ is the fiducial volume. In this form, we can see that it does not contradict the recipe below \eqref{semideform}, which obtains $\tilde{T}\tilde{\overline{T}}$ by  taking $\l \rightarrow \l e^{-2\Phi}$  and $V \rightarrow \tilde{V}$ in $T\overline{T}$, i.e. $\tilde{T} \tilde{\overline{T}}(\l, \tilde V) = T\overline{T}(\l e^{-2\Phi},\tilde V)$. As before we restrict to homogeneous $\Phi$. Consistency between these two equations requires $e^{4\Phi} T\overline{T}(\l,V) = T\overline{T}(\l e^{-2\Phi}, \tilde V)$. This is true due to dimensional analysis, which lets us write $T\overline{T}(\l, V) =\l^{-2} f(\l/V)$, $T\overline{T}(\l e^{-2\Phi}, \tilde V) = \l^{-2} e^{4\Phi} f(\l e^{-2\Phi}/\tilde V)$ and use $V = e^{2\Phi}\tilde V$. } 
Rewriting \eqref{eomhcons} in terms of the physical metric, we have
\be
-R+8\pi \mathfrak{b}^2\mu -8\pi \mathfrak{b}^2 \lambda T\overline{T}=0.
\ee
On the physical metric the $T\bar{T}$ operator is the one constructed by the bulk Brown-York stress tensor for Dirichlet boundary conditions, \eqref{tdbc}. Furthermore, using the dictionary for our proposal we have $\mu=(K\ell-2)/16\pi G\ell$, $\lambda=16\pi G\ell$, and $\mathfrak{b}^2=6/c_m=4G/\ell$. So the equation of motion reduces precisely to \eqref{traceeq}.

Another way to write the bulk Hamiltonian constraint, or the Liouville equation of motion, is as the trace relation 
\be\label{tracerelation}
\tilde T^\m_\m = - \f{c_m}{24\pi} (\tilde{R}-2 \tilde{\Box}\Phi) - 2\l \tilde T\tilde{\overline{T}}e^{-2\Phi}\,.
\ee


\subsection{Sphere partition function}\label{spherebdry}
In this section we will calculate the sphere partition function in our deformed theory and reproduce the bulk formula \eqref{sphereZ}, which we repeat here for convenience:
\be\label{sphereZ2}
\log Z[S^2] = \f{c_m}{6} \, \log \f{K\ell+2}{K\ell-2}\,.
\ee
First we will treat the case without perturbation, which effectively reduces to a calculation in timelike Liouville theory  \cite{Anninos:2021ene}. This should match the $K\ell \rightarrow 2$ limit of the above. We recall the action
\be
S_{tL} =\f{1}{4\pi \mathfrak{b}^2} \int d^2 x \sqrt{\tilde{g}} \left[-(\tilde{\nabla}\Phi)^2 - \tilde{R} \Phi + 4\pi \mathfrak{b}^2\m  e^{2\Phi}\right].
\ee
The saddle-point equation is given by
\be
2\tilde\Box \Phi-\tilde{R}  + 8\pi \mathfrak{b}^2 \m e^{2\Phi_\star} = 0 \,.
\ee
Focusing on constant $\Phi$ and a fiducial metric of a unit-radius $S^2$ with $\tilde{R}=2$, this becomes
\be
-2 + 8\pi \mathfrak{b}^2 \m e^{2\Phi_\star} = 0 \implies 2\Phi_\star = \log\f{1}{4\pi \mathfrak{b}^2 \m}+ 2\pi i n\,,\quad n \in \mathbb{Z}\,,
\ee
where we used $\tilde{R} = 2$ for the fiducial metric given by a unit-radius $S^2$. We will focus on the real saddle $n=0$ for now but will bring in the complex saddles once we do our nonperturbative calculation. Evaluating the Liouville action on this saddle gives
\be
\log Z_{tL} \approx \f{1}{\mathfrak{b}^2}\log \f{1}{4\pi \mathfrak{b}^2\m} - \f{1}{\mathfrak{b}^2}\,.
\ee
We now use $\mathfrak{b}^2=6/c_m$ and add in the matter sector to obtain the full action:
\be
\log Z = \log Z_{tL} + \log Z_m - \f{c_m}{6} \log \l = \log Z_m +  \f{c_m}{6} \log \f{4}{\m \l} +\f{c_m}{6}\log \f{c_m }{96\pi}- \f{c_m}{6}\,.
\ee
We split off the matter theory anomaly from $\log Z_m$, and we split off a particular $\m\l$-independent piece. There is a constant ambiguity in the matter sector, which we will take once and for all to cancel the $\m\l$-independent piece. Altogether this gives
\be
\log Z = \log Z_{tL} + \log Z_{m} = \f{c_m}{6} \log \f{4}{\m \l} = \f{c_m}{6}\log\f{4}{K\ell-2} 
\ee
where in the final expression we used our identification of parameters $\m = (K\ell-2)/\l$. This precisely matches \eqref{sphereZ2} as $K\ell\rightarrow 2$.

Now we compute the first order correction. In our Liouville description this requires adding to our matter theory a deformation $\l \tilde T\tilde{\overline{T}} e^{-2\Phi}$. So our action is given by \eqref{lineardef}, and its  saddle point equation is 
\be
-2 \tilde{\Box} \Phi = - \tilde{R} + 8 \pi \mathfrak{b}^2 \m e^{2\Phi} - 8\pi \mathfrak{b}^2 \l \tilde T\tilde{\overline{T}}e^{-2\Phi}\,.
\ee
Importantly the matter stress-tensor operator is calculated on the fiducial metric, a unit-radius sphere. The conformal anomaly in the matter theory $\tilde T^\m_\m =-c_m \tilde{R}/(24\pi)$ with $\tilde{R}=2$ means we have $\tilde T\tilde{\overline{T}}= - c_m^2/(2304\pi^2)$. Therefore  we have the constant-dilaton saddle point 
\be
2\Phi_\star = \log\,\f{1 + \sqrt{1 - \f{c_m^2 \mathfrak{b}^4 \l \m}{36 }}}{8 \pi \mathfrak{b}^2 \m}\approx \log \f{1}{4\pi \mathfrak{b}^2 \m} - \m\l \,\f{c_m^2 \mathfrak{b}^4}{144 } \,,
\ee
which we expanded to linear order in $\l$. Including the matter action, our full on-shell action to linear order is therefore 
\be
\log Z \approx \f{1}{\mathfrak{b}^2}   \log \f{4}{\m\l} + \m\l\, \f{c_m^2 \mathfrak{b}^2}{144} = \f{c_m}{6}\log\f{4}{K\ell-2} +\f{c_m}{24}(K\ell-2)\,,
\ee
where in the final expression we used $\mathfrak{b}^2 = 6/c_m $ and $\m = (K\ell-2)/\l$. This precisely reproduces the leading correction to \eqref{sphereZ2} as $K\ell \rightarrow 2$.

Now we do the calculation to all orders to reproduce \eqref{sphereZ2} exactly. 
The nonlinear deforming operator for the ordinary $T\overline{T}$ flow on $S^2$ has been worked out \cite{Donnelly:2018bef}. This is done by assuming a homogeneous stress tensor $\tilde T_{\m\n} = \a \tilde{g}_{\m\n}$ and plugging into the trace flow equation \eqref{tracerelation}, which for homogeneous $\Phi$ takes the form
\be\label{tracerelation}
\tilde T^\m_\m = - \f{c_m}{24\pi} \tilde{R} - 2\l \tilde T\tilde{\overline{T}}e^{-2\Phi}\,.
\ee
Solving this gives
\be\label{aeqn}
\a = \f{2}{\l e^{-2\Phi}} \left(1 - \sqrt{1 + \f{c_m \l e^{-2\Phi}}{24 \pi}}\right).
\ee
We find our dressed operator by calculating $\tilde T\tilde{\overline{T}}e^{-2\Phi}$ with $\tilde T\tilde{\overline{T}}= -\a^2/4$, which we then insert into our action:
\be\label{flowedaction}
S(\l) = \int d^2 x \sqrt{\tilde{g}}\int d\l \,e^{2\Phi}\f{-\left(1 - \sqrt{1 + \f{c_m\l e^{-2\Phi}}{24\pi}}\right)^2}{\l^2} + \f{c_m}{6} \log \l + S_{tL}+S_{m}\,.
\ee
As mentioned below \eqref{semideform}, this is the same deforming operator we would get by doing the standard irrelevant $T\overline{T}$ flow and then taking $\l \rightarrow \l e^{-2\Phi}$ at the end. Integrating this, we get
\be\label{logform}
S(\l) = \f{c_m}{6} \left(1-f(\l e^{-2\Phi}) + 2 \log f(\l e^{-2\Phi})\right) +\f{c_m}{6} \log \l +S_{tL} + S_m\,,
\ee
\be
 f(\l e^{-2\Phi}) = \f{2}{1 + \sqrt{1 + \f{c_m \l e^{-2\Phi}}{24\pi}}}\,.
\ee
We can also write this in the form
\begin{align}\label{arcsinhform}
S(\l) = -\f{c_m}{3} \sinh^{-1} \left(\sqrt{\f{24\pi}{c_m\l e^{-2\Phi}}}\right) - \f{8\pi }{\l e^{-2\Phi}} \left(\sqrt{\f{c_m \l e^{-2\Phi}}{24\pi} + 1}-1\right)\nn\\
\hspace{-10mm}+\,\f{c_m}{6}\left(1 + 2 \Phi + \log \f{96\pi}{c_m}\right) + S_{tL} + S_m\,.
\end{align}
 The saddle-point equation for $\Phi$  has solution
\be\label{phistarsphere}
2\Phi_\star = \log \f{c_m}{24 \pi \m (1 + \l \m/4)}\,.
\ee
This leads to an on-shell action
\be
\log Z[S^2] = \f{c_m}{6}\log\left(1 + \f{4}{\l\m}\right) = \f{c_m}{6} \log \f{K\ell+2}{K\ell-2}\,.
\ee
This precisely reproduces the bulk answer \eqref{sphereZ2}. The on-shell action comes entirely from the $\sinh^{-1}$ piece. As required by our correspondence, our saddle point \eqref{phistarsphere} satisfies $e^{2\Phi_\star} = r_c^2$, with $r_c$ the location of the boundary given in \eqref{keqnsphere}. 

Pure Liouville theory has a $PSL(2,C)$ symmetry on $S^2$ \cite{Zamolodchikov:1995aa}. This is just the conformal symmetry, under which the Liouville field $\Phi$ transforms in a funny way, since the primaries are the vertex operators. In the semiclassical limit this symmetry is manifest at the level of the action, and it extends to our deformed theory as well. This matches the $PSL(2,C)$ symmetry in the bulk, meaning we have a divergent Vol$(PSL(2,C)$ factor for the sphere partition function in both the bulk and boundary theories. 

Let's match the complex saddles as well. In the form \eqref{arcsinhform}, we can see the $\sinh^{-1}$ is multivalued and may belie additional saddles offset by $\pi i n$. These saddles are easy to see directly from the form \eqref{logform}. The saddle-point equation following from that action depends only on $e^{-2\Phi}$, meaning we have a family of complex saddles
\be
2\Phi_\star =  \log \f{c_m}{24 \pi \m (1 + \l \m/4)} + 2\pi i n\,,\qquad n \in \mathbb{Z}\,.
\ee
The action itself, however, has a term that is linear in $\Phi$ from $S_{tL}$. This leads to a different value for the on-shell action of these complex saddles which precisely reproduces \eqref{zcomplex}:
\be
Z[S^2] = \exp\left[ \f{c_m}{6} \log \f{K\ell+2}{K\ell-2} + \f{ i \pi n c_m}{3}\right]\,.
\ee
We can recover the flat-space sphere partition function by taking $\m \l \rightarrow \infty$ and identifying $\mu\l = K\ell-2 \approx K\ell$, which leads to 
\be
\text{Flat space limit:}\qquad\log Z[S^2] =\f{2c_m}{3\m\l}= \f{1}{GK} \,.
\ee
This flat-space limit does not require us to explicitly scale $c_m$, i.e. in the microscopic theory we just have the same holographic CFT deformed by $\tilde T\tilde{\overline{T}}$ and coupled to Liouville theory, but we simply scale the $\tilde T\tilde{\overline{T}}$ coupling large (in units of the cosmological constant $\m$).

\subsection{Torus partition function}\label{torusbdry}
Now we use our deformed timelike Liouville theory to calculate the torus partition function. Recall that the bulk answer we would like to reproduce is given by 
\be\label{logzfull2}
\log Z(\tilde{\b}, \tilde{\Omega}) =\begin{cases} \dfrac{\pi^2 \ell}{2G \tilde{\b} (1-\tilde{\Omega}^2) }\,\dfrac{K\ell - \sqrt{K^2 \ell^2-4}}{2} \,,\qquad \tilde{\b}^2(1-\tilde{\Omega}^2) < 4\pi^2 \\[1.5em]
\dfrac{\tilde{\b}\ell}{8 G} \, \dfrac{K \ell - \sqrt{K^2 \ell^2-4}}{2}\,,\qquad \qquad \qquad\tilde{\b}^2(1-\tilde{\Omega}^2) > 4\pi^2
\end{cases}.
\ee
We can see that the first factor in either answer is the usual free energy of the holographic CFT with central charge $c_m = 3\ell/2G$. Since $K\ell$ is fixed, we interpret the second factor as  modifying the central charge, $c_m \rightarrow c_{\text{eff}} = c_m (K\ell - \sqrt{K^2\ell^2-4})/2$. This square-root formula is the nontrivial formula to produce. We will begin by calculating the high-temperature answer. The same strategy can be used for the low-temperature answer. We will carry around the factors of $\tilde{\Omega}$, but one can also obtain the answer for $\tilde{\Omega} = 0$ and bring in the angular velocity through the thermal effective action \cite{Bhattacharyya:2007vs, Benjamin:2023qsc} or Lorentz invariance \cite{Shaghoulian:2015lcn, Allameh:2024qqp}. 

We take our physical metric to be $ds^2 = e^{2\Phi} d\tilde{s}^2$, where the fiducial metric is a flat torus, so $\tilde{R}=0$. Let's first calculate at linear order in  $\l$. Our fiducial metric is a twisted torus with $(\tilde{\t}, \phi) \sim (\tilde{\t}, \phi + 2\pi) \sim (\tilde{\t} + \tilde{\b}, \phi + i \tilde{\b}\tilde{\Omega})$. The saddle point equation is
 \be
-2 \tilde{\Box} \Phi = 8 \pi \mathfrak{b}^2 \m e^{2\Phi} - 8\pi \mathfrak{b}^2\l \tilde T\tilde{\overline{T}}e^{-2\Phi}\,.
 \ee
 As can be seen, in pure timelike Liouville theory (i.e. with $\l =0$) there is no semiclassical saddle point for constant dilaton. At first order in $\l$, however, we have the following constant-dilaton saddle:
 \be
e^{4\Phi} = \f{\l \tilde T\tilde{\overline{T}}}{\m}\,.
 \ee
 Our matter theory evaluated on the fiducial metric has
 \be
\log Z_m = \f{\pi^2 c_m}{3 \tilde{\b}(1-\tilde{\Omega}^2) }\,,\qquad \tilde T\tilde{\overline{T}} = \f{\pi^2c_m^2}{144 \tilde{\b}^4(1-\tilde{\Omega}^2)^2}\,.
 \ee
This is correct for an arbitrary CFT in a perturbative expansion around high temperature. The lack of any subleading perturbative terms in $\tilde{\b}$ follows from the thermal effective action. Nonperturbative corrections can also be eliminated by assuming a sparse spectrum of light states \cite{Hartman:2014oaa, Anous:2018hjh, Dey:2024nje} or a particular pattern of higher-form symmetry breaking \cite{Shaghoulian:2016xbx, Shaghoulian:2020omr}. Our saddle point becomes \label{units in log?}
\be
2\Phi_\star  = \log\left( \f{\pi c_m }{12 \tilde{\b}^2(1-\tilde{\Omega}^2)} \sqrt{\f{\l}{\m}}\right).
\ee
Adding in the contribution of the matter sector, our on-shell action becomes
\be
\log Z = \f{\pi^2 c_m}{3 \tilde{\b}(1-\tilde{\Omega}^2) } -\f{\pi ^2 c_m}{3 \tilde{\b}(1-\tilde{\Omega}^2)} \sqrt{\l\m}  =  \f{\pi^2 c_m}{3 \tilde{\b}(1-\tilde{\Omega}^2) }(1-\sqrt{K\ell-2})\,.
\ee
In the final expression we used our dictionary $\m = (K\ell-2)/\l$. This agrees precisely with the first correction to the high-temperature partition function in \eqref{logzfull2} in an expansion around $K\ell = 2$.

Now we work to all orders. The deforming operator is given in \eqref{ttbartildetorus}:
\be\label{ttbartorus}
\tilde T\tilde{\overline{T}}= \f{2 }{\l^2 e^{-4\Phi} }\left(\f{1-\tilde{\Omega}^2+ \pi c_m \l e^{-2\Phi} /(12\tilde{\b}^2)}{\sqrt{(1-\tilde{\Omega}^2)(1-\tilde{\Omega}^2 + \pi c_m \l e^{-2\Phi} /(6\tilde{\b}^2)})}-1\right).
\ee
We put this into our flow equation
\be
\f{\p S}{\p \l} = \int d^2 x \sqrt{\tilde{g}} \tilde T\tilde{\overline{T}} e^{-2\Phi}\,,
\ee
integrate, and add in the matter and timelike Liouville sectors to get our deformed theory. It has a saddle point given by
\be\label{phistar}
2\Phi_\star = \log \left(\f{\pi c_m \l}{12\tilde{\b^2}(1-\tilde{\Omega}^2)}\,\f{2 + \l \m -\sqrt{\l \m(4 + \l \m)}}{\sqrt{\l \m(4 + \l \m)}}\right).
\ee
This saddle can also be more directly obtained from \eqref{lioueom}. Evaluating our action on this saddle point gives
\be\label{torusfinal}
\log Z = \f{\pi^2 c_m}{6 \tilde{\b}(1-\tilde{\Omega}^2)}\left(2 + \l \m - \sqrt{\l \m(4 + \l \m)}\right) = \f{\pi^2 c_m}{3\tilde{\b}(1-\tilde{\Omega}^2)} \f{K\ell - \sqrt{K^2 \ell^2-4}}{2}\,,
\ee
where in the final expression we used our dictionary $\m = (K\ell-2)/\l$. As in the calculation of $\log Z[S^2]$, our saddle point \eqref{phistar} satisfies $e^{2\Phi_\star} = r_c^2$, with $r_c$ the location of the boundary given in \eqref{ranswers}. The low-temperature partition function can be derived in a completely analogous manner, or by the modular transformation $\tilde\b \rightarrow 4\pi^2/(\tilde \b (1-\tilde \Omega^2))$.

As before, we access the flat-space limit as $\m \l \rightarrow \infty$:
\be
\text{Flat space limit:}\qquad \log Z(\tilde{\b}, \tilde{\Omega}) =\begin{cases} \dfrac{\pi^2}{2 \tilde{\b} (1-\tilde{\Omega}^2)G K } \,,\qquad \tilde{\b}^2(1-\tilde{\Omega}^2) < 4\pi^2 \\[1.5em]
\dfrac{\tilde{\b}}{8 G K} \,,\qquad \qquad \qquad\,\tilde{\b}^2(1-\tilde{\Omega}^2) > 4\pi^2
\end{cases}.
\ee
We see that using $c_m = 3\ell/(2G)$ leads to the factors of $\ell$ dropping out when translating into a bulk expression. This extends the modular-invariant structure of 3d flat space discussed in \cite{Banihashemi:2024yye}  to include an angular potential. The effective central charge is identified as
\be
c_{\text{eff}}^{\text{flat}} = \f{3}{2GK}\,.
\ee
We will comment further on the flat-space limit in the discussion section.
\section{Stress tensor correlation functions}\label{correlators}
A simple check we can provide on our proposal is the calculation of stress tensor correlation functions. Since our dual theory is claimed to be a two-dimensional CFT with vanishing anomaly central charge, this means that the stress tensor is null and all its correlators with itself on the sphere need to vanish. We will check this boundary prediction with a bulk calculation. We use the formula 
\begin{align}
\hspace{-5mm}\langle T_{\m\n}(x)\rangle_{g+h} &= \int [D\Phi] e^{-S + \f 1 2 \int d^d y \sqrt{g(y)} T_{\a\b}(y)h^{\a\b}(y)}T_{\m\n}(x)\nn\\
&\approx \langle T_{\m\n}(x)\rangle_g + \f 1 2\int d^d y\sqrt{g(y)} \langle T_{\m\n}(x) T_{\a\b}(y)\rangle_g h^{\a\b}(y)\nn \\
&\hspace{16mm}+\f 1 8 \int d^d y d^d z \sqrt{g(y)g(z)} \langle T_{\m\n}(x) T_{\a\b}(y) T_{\g\d}(z)\rangle_g h^{\a\b}(y) h^{\g\d}(z) + \cdots\,, \label{Tseries}
\end{align}
where $g_{\m\n}$ refers to the metric on the sphere and $h_{\m\n}$ is the source. We know that $\langle T_{\m\n}\rangle_g = 0$, both from a direct bulk calculation and from the assumption of having a conformal field theory with vanishing conformal anomaly. The relevant stress tensor calculated from the bulk is the conformal Brown-York stress tensor.

 We can split up an arbitrary metric perturbation into a Weyl mode and a ``shape" mode. By the uniformization theorem in 2d, any such shape mode on $S^2$ is pure gauge, and we will therefore ignore it. (This statement is sufficient to show the vanishing of correlators when we restrict the source $h_{\m\n}$ to a shape deformation.) We source the boundary Weyl mode as $ds^2 = e^{2\Phi(z,\bar{z})} d\Omega_2^2$. A general metric ansatz for the bulk solution preserving reflection symmetries and $S^2$ topology is given by
\be
ds^2 = e^{2\phi(\rho, z, \bar{z})} d\Omega_2^2 + e^{2f(\rho, z, \bar{z})} d\rho^2\,,\qquad \phi(\rho_c, z, \bar{z}) = \Phi(z, \bar{z})\,.
\ee
The fields $f$ and $\phi$ are allowed to depend on all three coordinates. Now we simply calculate the conformal Brown-York stress tensor \eqref{cbcstress1} on this background and find that it vanishes. So we get $\langle T_{\m\n}\rangle_{g+h} = 0$, indicating all correlators on the sphere vanish. The fact that this one-point function vanishes is consistent with the boundary expectation: since the boundary geometry is Weyl-equivalent to the sphere, and we have no Weyl anomaly, we should get the same stress tensor expectation value as on the sphere. This argument can be checked for some low-point functions \cite{allameh2} along the lines of  \cite{Kraus:2018xrn}.

\section{Discussion}\label{discsec}
In this paper we have proposed a two-dimensional conformal field theory dual to AdS$_3$ gravity with conformal boundary conditions. Even though the gravitational theory generically lives at finite cutoff, the dual theory -- described by an exactly marginal deformation of timelike Liouville theory coupled to the holographic CFT$_2$ dual to AdS$_3$ -- exhibits extensivity in its high-temperature entropy. We provided some checks on the theory in the semiclassical limit, where we were able to match the torus and sphere partition functions and reproduce the bulk Hamiltonian constraint through the Liouville field's equation of motion. These calculations let us reproduce the effective central charge \cite{Anninos:2024wpy, Banihashemi:2024yye}
\be
c_{\text{eff}} = \f{3\ell}{2G}\f{K\ell - \sqrt{K^2\ell^2-4}}{2}
\ee
controlling the black hole entropy. 
This central charge decreases along the flow even though the deformation is by an irrelevant operator (dressed to become marginal).

In the proposed duality, the Liouville field is related to the radial direction in the bulk, with its saddle point value providing the location of the wall in our bulk, $e^{2\Phi_\star} = r_c^2$ for the gauge chosen in this paper. This is different than the general picture advocated by Polyakov \cite{Polyakov:1998ju} or the AdS$_3$ picture advocated by Coussaert, Henneaux, and van Driel \cite{Coussaert:1995zp}. Shifts in the saddle point-value $\Phi_\star$, which in Liouville theory correspond to rescaling the physical cutoff, map to a rescaling of the bulk cutoff $r_c$. In the sphere and torus partition functions, decreasing $K\ell$ moved the bulk wall outward, allowing more of spacetime to be accessed. In Liouville theory decreasing $K\ell$ corresponds to decreasing the cosmological constant, and therefore the strength of the potential. This moves the Liouville wall outward. In the torus partition function, decreasing $\tilde{\b}$ or increasing $\tilde{\Omega}$ also move the bulk wall outward. These correspond to increasing the energy of the state, which also allows more of the spacetime to be explored. This flexibility of conformal boundary conditions is essential to highlight: at fixed $K\ell$, which corresponds to a fixed theory, higher energy states explore more of the spacetime, with no upper limit on the energy or the amount of spacetime that can be uncovered. This is reflected in the dynamics of the Liouville wall in our proposed dual theory. 


There are some interesting extensions that would be worth pursuing. A straightforward one would be to study the addition of matter fields to the bulk. In that case our dual Liouville theory has to be deformed by a more general operator. The recipe to write down these operators can be found in \cite{Hartman:2018tkw}. This would also allow a computation of correlation functions beyond the stress-tensor sector, which will help characterize the degree of (non)locality of our theory. While the high-temperature thermodynamics scales extensively and therefore suggests locality, it is not clear that this will be preserved at the level of correlation functions. (At high temperature, the cutoff goes to infinity, so the locality may be a mirage due to this effect.) In particular, the integrated $\tilde{T} \tilde{\overline{T}}$ operator leads to higher derivative terms that generically spoil locality. Another straightforward extension one would be to study our dual theory on a negatively curved fiducial metric. Pure timelike Liouville theory only has a real saddle for a positively curved fiducial metric, and requires complex saddles to accommodate a negatively curved fiducial metric. Our deformation, however, introduces a term that can be offset against the cosmological constant term to find a real saddle, in the same way that we found a saddle point for a flat fiducial metric (i.e. the torus calculation in Section \ref{torusbdry}). Two important checks on our proposal would be to consider saddles with non-homogeneous $\Phi$ and to compute loop effects.

We should comment on a somewhat peculiar feature of our proposed duality. In the usual AdS$_3$/CFT$_2$ duality, the vacuum of the CFT$_2$ on the plane is represented by the Poincar\'e patch geometry in the bulk. Its stress tensor vanishes. It conformally transforms to the ground state on the cylinder, represented by global AdS$_3$, which has nonvanishing stress tensor due to the conformal anomaly. In our proposed duality, however, the anomaly vanishes. We nevertheless have a vanishing stress tensor for the Poincar\'e patch geometry and a nonvanishing stress tensor in global AdS$_3$, which seems inconsistent with a vanishing anomaly. The way to think about this is the following. Take a given geometry in the bulk corresponding to a state of the CFT$_2$ on a circle. To obtain the geometry corresponding to the conformal transformation of this state to the plane, we simply unwrap the spatial circle. This procedure assures that the stress tensor is the same between the plane and the cylinder. As far as the ground state is concerned, it implies that the state on the plane of minimal stress energy is given by the geometry
\be
ds^2 = -(1 +r^2/\ell^2)dt^2 + \f{dr^2}{1+r^2/\ell^2} + r^2 dx^2 \,,\qquad x \in \mathbb{R}
\ee
obtained by decompactifying the spatial circle of global AdS$_3$. Instead of mapping to the Poincar\'e patch we get a geometry that is singular as $r\rightarrow 0$. To get the Poincar\'e patch, we instead need to start with the $M=0$ BTZ black hole.

\subsection*{Peering toward the black hole singularity}
The $r=0$ surface of the rotating BTZ black hole is not a genuine curvature or conical singularity in pure gravity; it instead demarcates the region $r<0$ in which we have closed timelike curves. Generic matter couplings, however, will turn $r=0$ into a genuine curvature singularity \cite{Banados:1992gq}. Pushing the cutoff surface toward $r=0$ corresponds to taking the limit $K\ell \rightarrow \infty$, which in our boundary dual corresponds to $\m \l \rightarrow \infty$. Nothing particularly pathological happens in this limit; we will comment in the next subsection on an interpretation of this limit in terms of 3d flat-space physics. 

We can also try to peer toward the singularity. This is done by considering the patch $r \in [0,r_c]$ with $r_c < r_-$. (We assume we have some matter which leads to a curvature singularity at $r=0$ and we therefore allow spacetime to end there.) This flips the sign of $K$, which in our dual description corresponds to $\m < 0$. For an ordinary real contour for $\Phi$ as considered in spacelike Liouville theory, this leads to an unbounded potential, so it is not usually considered. The relevant contour in timelike Liouville theory has to be modified due to the already-existing negative-definite kinetic mode, so it is not clear whether $\m < 0$ is sensible. In fact, in this case the action can be thought of as minus the spacelike Liouville action.

While physically unmotivated, one could try to entertain going past $r = 0$ in pure gravity where there is no singularity. In other words, our flipped region $r \in [0, r_c]$ would not have to end at $r=0$. In that case, however, the geometry does not end and goes all the way to $r=-\infty$, reaching another AdS boundary. The existence of this extraneous boundary conflicts with our boundary value problem, which gives another reason to not include such a patch.

In the higher-dimensional cases considered in \cite{Banihashemi:2025qqi}, it is interesting that going deep inside the inner horizon of a black hole, while maintaining a patch $r \in [r_-, r_c]$, eventually led to $K < 0$. In AdS$_3$ the only way to obtain $K < 0$ is to flip the patch to include the singularity as considered above.

\subsection*{Flat space limit}

\begin{figure}
\centering
\begin{tikzpicture}[scale=1.4]

\draw[draw=none,pattern = crosshatch,pattern color=penrosered] (1,1) .. controls (0.45,0.25) and (0.45,-0.25) .. (1,-1) to (0,0) -- cycle;

\draw[-,penrosered!80!black,dashed,very thick] (-1,-1) to (1,1);
\draw[-,penrosered!80!black,dashed,very thick] (-1,1) to (1,-1);

\draw[-] (-2,0) to (0,-2) to (2,0);
\draw[-] (-2,0) to (0,2) to (2,0);

\node[penrosered!80!black] at (0.35,0.7) {\large$r_h$};
\node[penrosered!80!black] at (0.35,-0.7) {\large$r_h$};

\draw[-] (1,1) .. controls (0.45,0.25) and (0.45,-0.25) .. (1,-1);

\node at (0.775,0) {\large$r_c$};
\end{tikzpicture}
\hspace{40mm}
\begin{tikzpicture}[scale=1.4]

\draw[draw=none,pattern = crosshatch,pattern color=penroseblue] (1,1) .. controls (0.45,0.25) and (0.45,-0.25) .. (1,-1) to (0,0) -- cycle;

\draw[-,penroseblue!80!black,dashed,very thick] (-1,-1) to (1,1);
\draw[-,penroseblue!80!black,dashed,very thick] (-1,1) to (1,-1);

\draw[-] (-1,-1) to (0,-2) to (1,-1);
\draw[-] (-1,1) to (0,2) to (1,1);

\draw[-,decorate,decoration={snake,amplitude=0.5mm,segment length=3.8mm}] (1,-1) to (1,1);
\draw[-,decorate,decoration={snake,amplitude=0.5mm,segment length=3.8mm}] (-1,1) to (-1,-1);

\node[penroseblue!80!black] at (0.35,0.7) {\large$r_h$};
\node[penroseblue!80!black] at (0.35,-0.7) {\large$r_h$};

\draw[-] (1,1) .. controls (0.45,0.25) and (0.45,-0.25) .. (1,-1);

\node at (0.775,0) {\large$r_c$};
\end{tikzpicture}

\caption{(a) The Penrose diagram for Rindler space, the flat-space solution with horizon at finite temperature and vanishing angular velocity. The relevant patch (shown in red) is obtained by the flat-space limit $K\ell \rightarrow \infty$ of the non-rotating BTZ black hole, which zooms into the outer horizon. (b) The Penrose diagram for the flat-space solution with horizon at finite temperature and nonzero angular velocity. The relevant patch (shown in blue) is a remnant of the region beyond the inner horizon of the rotating BTZ black hole. By analytically continuing this region beyond the horizons after the flat-space limit, one recovers the null asymptotics. The same Penrose diagram governs hyperbolic and planar horizons in higher-dimensional flat space \cite{Banihashemi:2025qqi}.}
\label{PenroseFSC}
\end{figure}
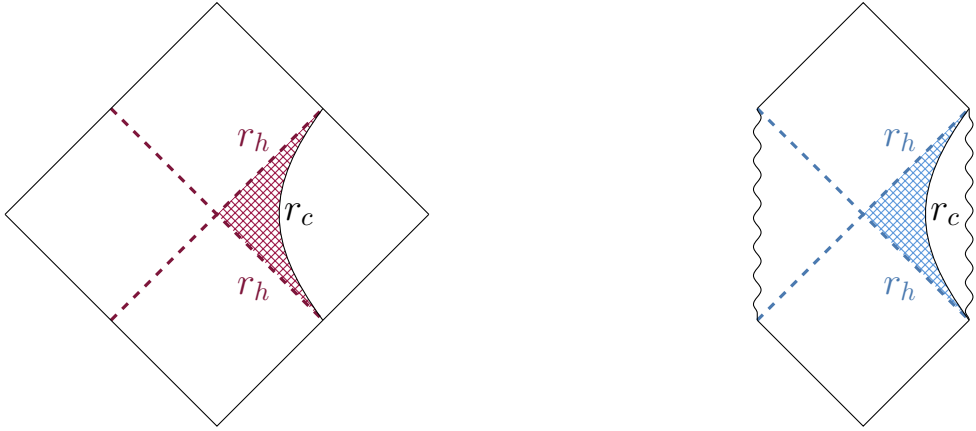

In the bulk calculations of the sphere and torus partition functions, the flat-space limit was accessible through the limit $K \ell \rightarrow \infty$. An interesting aspect of the dual Liouville theory is that one can take this flat-space limit without scaling the central charge of the holographic CFT. Since our dictionary identifies the cosmological constant term in the Liouville coupling as $\m = (K\ell-2)/\l$, we can take the flat-space limit by scaling $\m \l \rightarrow \infty$. In our boundary calculation of the sphere and torus partition functions, this limit precisely reproduced the flat-space answers. Thus our theory for 3d flat space is governed by
\be
c = 0\,,\qquad c_{\text{eff}}= \f{3}{2GK}\,.
\ee
We can use this limit to connect to some work on holography in 3d flat space, see e.g. \cite{Barnich:2006av, Barnich:2012aw, Bagchi:2012cy, Bagchi:2012xr, Bagchi:2013lma, Detournay:2014fva, Bagchi:2016bcd}. It will help to first discuss what happens to the BTZ geometry and thermal AdS in the flat-space limit $K\ell \rightarrow \infty$. We first consider the non-rotating BTZ geometry, corresponding to $\tilde{\Omega} = 0$. The flat-space limit $K\ell\rightarrow \infty$ pushes us toward the BTZ horizon, giving us a piece of the Rindler geometry. The relevant patch and its analytic continuation beyond the horizon in 3d flat space is presented in Figure \ref{PenroseFSC}(a).

Now we consider the rotating BTZ black hole, corresponding to $\tilde{\Omega} \neq 0$. Generically, for black holes with inner and outer horizons, the outer horizon scales with $\ell$ and goes off to infinity in the flat space limit, whereas the inner horizon remains finite. In the flat-space limit $K\ell \rightarrow \infty$ our cutoff surface gets pushed deep inside the inner horizon. This gives a solution of 3d flat space gravity, sometimes called a ``flat space cosmology" in the literature. The relevant patch and its analytic continuation beyond the horizon in 3d flat space is presented in Figure \ref{PenroseFSC}(b). The Penrose diagram is the same as that of the hyperbolic and planar horizons in higher-dimensional flat space (see Figure 4 of \cite{Banihashemi:2025qqi}).

Beyond these two solutions, we also have thermal flat space $ds^2 = d\t^2 + dr^2 + r^2 d\phi^2$, with $(\t, \phi) \sim (\t + \tilde{\b}, \phi + i \tilde{\b}\tilde{\Omega})$ and $\phi \sim \phi + 2\pi$. This is obtained by a flat-space limit of thermal AdS. The transition between this vacuum geometry and the horizonful geometries remains at the modular-invariant locus $\tilde{\b}^2(1-\tilde{\Omega^2}) = 4\pi^2$.

Much of the work on the flat-space limit of AdS$_3$ has tried to characterize the dual theory in terms of a system with a Galilean conformal algebra, which is isomorphic to BMS$_3$ \cite{Barnich:2006av}. (The work also often considers Dirichlet boundary conditions, although see \cite{Detournay:2014fva} for an analysis with the same boundary action we have used here.) BMS$_3$ represents the asymptotic symmetries of 3d flat space. This algebra can be thought of as the $\ell\rightarrow \infty$ limit of the asymptotic symmetry algebra of AdS$_3$. This is formalized in terms of the In\"on\"u-Wigner contraction of two copies of the Virasoro algebra. Our theory, however, remains a relativistic conformal field theory. This leads to various differences, for example our transition point between the vacuum solution and the horizonful solution occurs at the usual modular-invariant locus, which differs from the transition found in \cite{Bagchi:2013lma}.   Interestingly, the Galilean conformal algebra has one copy of Virasoro with $c=0$. To connect to approaches using the asymptotic boundary, one can supplement the limit of the flow taken here with another flow, along the lines of \cite{Gorbenko:2018oov, Coleman:2021nor}. We will discuss an implementation of this two-part flow for de Sitter spacetime in the next subsection.

\subsection*{de Sitter limit}
In our proposal thus far, three-dimensional flat space and AdS$_3$ can be considered by different couplings of the Liouville sector to the same matter theory. It is natural to ask whether dS$_3$ can be considered in a similar fashion. The bulk theory is now Einstein gravity with positive cosmological constant. The sphere and torus partition functions of a putative boundary dual have been calculated in this theory with conformal boundary conditions. They are given by \cite{Anninos:2024wpy}
\be\label{ds3}
\log Z[S^2] = \f{ic_m}{6} \log \f{K\ell - 2 i}{K\ell + 2i}\,,\qquad \log Z[\mathbb{T}^2] = \f{\pi^2 c_m}{3 \tilde{\b}(1-\tilde{\Omega}^2)} \f{-K\ell + \sqrt{K^2\ell^2 + 4}}{2}\,,
\ee
where $c_m=3\ell/2G$ and $\ell$ is now the de Sitter radius. The sphere partition function is real even though it looks complex, as it can be written as $\log Z[S^2] = \f{c_m} 3 \tan^{-1}\f{2}{K\ell}$. It is written the above way to indicate the similarity to the sphere partition function in AdS$_3$. These formulas provide a natural target to reproduce from a putative boundary theory.

 At a formal level, capturing dS from AdS can be attempted via $\ell \rightarrow i\ell$. Many souls have perished here, although see \cite{Anninos:2011ui, Anninos:2017eib}. In our dual theory this would require $c_m \rightarrow i c_m$ in addition to other complexifications. It is not known how to make sense of such a procedure, and in any case this would change our matter theory. We expect the answers \eqref{ds3} to come, at a technical level, mostly from the coupling to Liouville theory, as we saw in the AdS$_3$ case. So perhaps we can complexify just the Liouville theory. Formally, for semiclassical states with $T\overline{T} \sim O(G^{-2})$, we can take the original matter theory and couple it to a complexified Liouville theory as follows:
 \be
S_{Liou} = \f{-i}{4\pi \mathfrak{b}^2}\int d^2 x \sqrt{\tilde{g}} \left(-(\tilde{\nabla}\Phi)^2 - \tilde{R}\Phi -4\pi \mathfrak{b}^2 i \m e^{2\Phi} -4\pi \mathfrak{b}^2\int d\l \,\tilde T\tilde{\overline{T}} e^{-2\Phi}\right)
 \ee
with parameters
\be
\m = \f{K\ell-2 i}{\l}\,,\qquad \l = 16 \pi G \ell\,,\qquad \f{1}{\mathfrak{b^2}} = \f{c_m}{6} = \f{\ell}{4G}\,.
\ee
Up to an innocuous constant piece, this coupling reproduces the dS$_3$ sphere and torus partition functions above by the same procedures as before. However, it is not clear that coupling this theory to the original matter CFT makes sense beyond the semiclassical saddles we have investigated.

Another idea is to use the massive deformation to the  AdS$_3$ matter theory given in  \cite{Gorbenko:2018oov}. We will also perform a two-part flow, versions of which are discussed in \cite{Gorbenko:2018oov, Coleman:2021nor}. In the first part of the flow, we perturb by the gravitationally dressed $\tilde T\tilde{\overline{T}}$ operator as done in this paper, and take $\m \l \rightarrow \infty$. The next part of the flow is defined by
\be
\f{\p S}{\p \l} =\int d^2 x \sqrt{\tilde{g}}\, \tilde T\tilde{\overline{T}}e^{-2\Phi}  +\f{c_m}{48\pi \l} \int d^2 x \sqrt{\tilde{g}}\,\tilde R- \f{2}{\l^2}\int d^2 x \sqrt{\tilde{g}}e^{2\Phi} \,.
\ee
This modified flow adds a $\l$-dependent contribution to the cosmological constant term in Liouville theory. 

To show how this works, let us recover $\log Z[S^2]$ in dS$_3$, written above. The matter theory flow is governed by the modified trace relation
\be
\tilde T^\m_\m = -\f{c_m}{24\pi}\,\tilde R - 2 \l \tilde T\tilde{\overline{T}} e^{-2\Phi} + \f{4}{\l}e^{2\Phi}\,.
\ee
Assuming that $\tilde T_{\m\n} = \a \tilde{g}_{\m\n}$ and plugging into this trace equation gives
\be
\a = \f{2}{\l e^{-2\Phi}} \left(1- \sqrt{-1 + \f{c_m\l e^{-2\Phi}}{24\pi}}\right),
\ee
with $\tilde T\tilde{\overline{T}}= -\a^2/4$. Our full action is therefore
\begin{align}
S(\l) = -\f{c_m}{3} \sin^{-1} \left(\sqrt{\f{24\pi}{c_m\l e^{-2\Phi}}}\right) - \f{8\pi }{\l e^{-2\Phi}} \left(\sqrt{\f{c_m \l e^{-2\Phi}}{24\pi} - 1}-1\right)\nonumber\\+\,\f{c_m} 6\left(1 + 2 \Phi + \log \f{96\pi}{c_m}\right) +S_{tL} + S_m\,,
\end{align}
When integrating $\tilde T\tilde{\overline{T}}$, we made a choice of integration constant so that the $\l \rightarrow \infty$ limit of the result agrees with the $\l \rightarrow \infty$ limit of the AdS$_3$ flow. This corresponds to the terms $(c_m/6)(1+ 2\Phi + \log (96\pi/c_m))$. This integration ``constant" actually includes the term $c_m \Phi/3$ explicitly written above. As a result it is more reasonable to include this term as part of the deformed Liouville action, where it simply cancels the usual $-c_m\Phi \tilde{R}/6$ term. In any case, using this rule leads to a saddle point 
\be
2\Phi_\star = \log \f{c_m\l}{6\pi(8 + \l \m(4 + \l \m))}\,.
\ee
Upon using $\m = (K\ell -2)/\l$,  this leads to an on-shell action
\be
\log Z = \f{i c_m}{6} \log \f{K\ell -2 i}{K\ell+2i}\,,
\ee
reproducing the sphere partition function above.

Another way to try to connect to dS$_3$ is as follows. We consider the bulk theory to be Einstein gravity with negative cosmological constant, as done throughout this paper. If we allow complex solutions, then a particularly simple one  is the ``negative dS$_3$" metric:
\be
\f{ds^2}{\ell^2} = -d\theta^2 - \sin^2 \theta d\Omega_2^2 = -(1-r^2/\ell^2)d\t^2 - \f{dr^2}{1-r^2/\ell^2}-r^2 d\phi^2\,.
\ee
We have written the metrics in spherical and toroidal slicing, respectively. These have an overall minus sign out front and therefore solve the bulk equations of motion with negative cosmological constant. (This is similar to the ``negative AdS$_2$" metrics discussed as solutions of gravity with a positive cosmological constant \cite{Maldacena:2019cbz}.) As long as the conformal class of metrics on the boundary allows overall minus signs, then this sits in the same $S^2$ family. 

\subsection*{Neumann boundary conditions and bc ghosts}
Our boundary conditions fix the conformal metric and the trace of the extrinsic curvature. This means we only integrate over the conformal mode of the boundary metric, as opposed to the full boundary metric. This is why there was no mention of $bc$ ghosts in our dual description, which is therefore not a non-critical string theory. As a result, the sphere partition function exhibits a divergence due to a Vol$(PSL(2,C))$ factor. This is analogous to the procedure in nearly AdS$_2$ spacetimes, where one has a $PSL(2,R)$ symmetry that is quotiented out \cite{Maldacena:2016upp}.

Of course, it is very natural to consider integrating over the full boundary metric. In the bulk this would correspond to Neumann boundary conditions. In the boundary theory it would introduce the $bc$ ghosts. In the traditional non-critical string theory, these $bc$ ghosts provide an inverse factor of Vol$(PSL(2,C))$, which renders the sphere path integral finite (see \cite{Mahajan:2021nsd} and \cite{ Anninos:2021ene} for some recent analyses in the spacelike and timelike Liouville theory, respectively). In this theory the total stress tensor becomes BRST exact and decouples from the physical spectrum. The calculations in the main text of this paper can be interpreted as calculations in a pre-Hilbert space which has yet to quotient out states equivalent by diffeomorphisms. 


\subsection*{Higher and lower dimensions}
For conformal boundary conditions in AdS$_{d+1}$, we have an action 
\be
I= -\f{1}{16\pi G} \int d^{d+1}x \sqrt{\mathcal{G}} \,(R+d(d-1)/\ell^2) - \f{1}{8\pi Gd} \int d^d x \sqrt{g} \,K\,.\label{cbcacthigher}
\ee
For a flat boundary, this action leads to totally finite answers, although there are uncured divergences when the boundary theory is placed on a curved background. (To analyze those situations, it might be worth switching to an ensemble where we fix the trace of the renormalized Brown-York stress tensor \cite{Parvizi:2025shq, Parvizi:2025wsg}.) This action leads to a conformal Brown-York stress tensor 
\be
8\pi G T^{CBC}_{\m\n} = K_{\m\n} - \f 1 d K g_{\m\n}
\ee
whose trace manifestly vanishes independent of background. The high-temperature partition function on an arbitrary background is extensive with a coefficient that is modified from the usual AdS/CFT coefficient by a factor $(K\ell/d - \sqrt{K^2\ell^2/d^2-1})^{d-1}$. Restricting to a flat boundary, we have a simple transform between ensembles 
    \be\label{mastertransformhigher}
    Z_{CBC}[\tilde{g}_{\m\n}, K] = \int dV Z_{DBC}[V]\,\exp\left(-\f{d-1}{d}\f{K\ell-d}{8\pi G\ell} \,V\right)\,.
    \ee
Thus we see that the higher-dimensional calculations have a broadly similar structure to the AdS$_3$ calculations, at least for flat boundaries. As we have analyzed here, the matter CFT needs to be deformed by the higher-dimensional version of the $T\overline{T}$ operator \cite{Taylor:2018xcy, Hartman:2018tkw}. Working in the semiclassical limit and writing the physical metric as $ds^2 = e^{2\Phi}d\tilde{s}^2$ means we can dress this operator as $\int \sqrt{\tilde{g}} (\tilde T_{\m\n}\tilde T^{\m\n} - \f 1 d (\tilde T^\m_\m)^2) e^{-d\Phi}$. 

We can also go to one lower dimension. As discussed in Appendix E of \cite{Banihashemi:2025qqi}, the natural thing to do in AdS$_2$ is to dimensionally reduce the conformal boundary conditions in AdS$_3$ or higher-dimensional AdS. This does not mean we fix the conformal class of metric and $K$ on the 1d boundary (where fixing the conformal class of metric in 1d does not fix anything). In the case of dimensionally reducing from AdS$_3$ in the $s$-wave sector, this fixes $K_{1d} + \p_n \log \phi$ and $\b_p/\phi$ at the boundary, where $\phi$ is the dilaton, $\p_n$ is a normal derivative to the boundary, $\b_p$ is the inverse proper temperature at the boundary, and $K_{1d}$ is the extrinsic curvature of the 1d boundary. Notice these two expressions are simply $K_{2d}$ and the inverse conformal temperature in the AdS$_3$ picture, respectively. The dimensional reduction of our Liouville theory gives Liouville quantum mechanics, also known as the minisuperspace approximation to Liouville theory. This is closely related to the Schwarzian theory, and it would be interesting to analyze the precise connection between the two theories for $K_{2d} > 2/\ell$ using \cite{Gross:2019ach}. The inverse factor of $PSL(2,C)$ we would obtain by integrating over the full boundary metric in AdS$_3$ with $S^2$ slices would reduce to an inverse factor of $PSL(2,R)$, giving a local way to obtain finite answers as advocated in \cite{Anninos:2021ydw}.

\subsection*{Fixed area quantities}
Fixed area quantities in Liouville theory are defined by inserting a delta function into the relevant path integral. For example, correlators of vertex operators are defined as \cite{Distler:1988jt}
\be
\big \langle \prod_i e^{\alpha_i \Phi}\big\rangle_A = \int D\Phi\, e^{-S_{L}} \prod_i e^{\alpha_i \Phi}\d\left(\int d^2 x \sqrt{g} - A\right).
\ee
To go back to fixed cosmological constant, we simply calculate
\be
\big\langle \prod_i e^{\alpha_i \Phi}\big\rangle_\m = \int dA \big \langle \prod_i e^{\alpha_i \Phi}\big\rangle_A
\ee
One of the motivations for considering fixed-area quantities  \cite{Seiberg:1990eb} is to deal with the divergence that occurs in the $\Phi \rightarrow -\infty$ part of field space. In the case of spacelike Liouville theory, such divergences will occur in the calculation of correlators
\be
\big\langle \prod_i e^{\alpha_i \Phi}\big\rangle_\m =  \int D\Phi\, e^{-S_{L}} \prod_i e^{\alpha_i \Phi}
\ee
unless 
\be
\sum_i \a_i > \f{c_L}{6}(2-2g)\,,
\ee
for $g$ the genus of the manifold. The $\Phi \rightarrow +\infty$ part of field space is tame due to the $\m e^{2\Phi}$ term. In perturbation theory it seems our deformation, if applied to spacelike Liouville theory, would help with convergence, e.g. at first order we have a term $\l \tilde T\tilde{\overline{T}} e^{-2\Phi}$. However, we know our full deformation introduces an effective coupling $\l e^{-2\Phi}$, so must be considered nonperturbatively in the $\Phi \rightarrow -\infty$ limit. In fact this seems to push us into the flat space limit.

An interesting aspect of our theory is that observables in $T\overline{T}$-deformed theories \cite{Zamolodchikov:2004ce, Smirnov:2016lqw, Cavaglia:2016oda, McGough:2016lol} are somewhat like fixed-area observables of our theory. To see this, let's focus on the partition function. We write  \eqref{mastertransform} and its inverse, with some slight change in notation $V \rightarrow A$ and using our dictionary $\m = (K\ell-2)/(16\pi G \ell)$, as 
\be
    Z_{CBC}[\m] = \int_0^\infty dA \,Z_{DBC}[A]e^{-\m A}
\,,\qquad Z_{DBC}[A] = \f{1}{2\pi i}\int_{\g-i\infty}^{\g + i\infty} d\m \,Z_{CBC}[\m] e^{+\m A}\,.
\ee
Up to the irrelevant factor $e^{-\m A}$, which is sometimes included in the definition of fixed-area states to get nicer scaling of the partition function \cite{Seiberg:1990eb}, these are precisely the transforms between fixed area and fixed cosmological constant. This is pretty manifest in the left transform. In the right transform, we simply note that the integral over $\m$ imposes the fixed-area constraint, since the only $\m$-dependent term in $S_{CBC}[\m]$ is $\m\int d^2 x \sqrt{\tilde{g}} \,e^{2\Phi}$. These transforms, however, are written for spacetime-independent $K$, $\Phi$. More generally the Dirichlet problem is not just fixing the total area, but is instead fixing the metric locally; this is reflected by the integrals above becoming path integrals.  

As a simple example of the integral transforms above, one can calculate the fixed-area partition function on the torus by plugging our $Z_{CBC}[\m]$ from \eqref{torusfinal} into the above transform. Evaluating by saddle point gives the torus partition function in $T\overline{T}$-deformed theories, which is reproduced in Appendix \ref{app:ttbartorus}. The sphere partition function can be reproduced similarly, and in the $A \rightarrow \infty$ limit gives the asymptotic AdS/CFT answer 
\be
\log Z_{DBC}[S^2; A \rightarrow \infty] = \f{c_m}{6} \log A\,.
\ee

\section*{Acknowledgments}
The authors would like to thank Dionysios Anninos for many illuminating conversations about these topics. The authors would also like to thank Batoul Banihashemi, Thomas Hartman, Raghu Mahajan, Beatrix M\"uhlmann, Sanjit Shashi, Vasudev Shyam, and Eva Silverstein for fruitful conversations. The authors are supported by DOE grant DE-SC001010.

\appendix

\section{Defining the theory at finite $c_m$}\label{app:qtm}
In this appendix we discuss the quantum theory at finite $c_m$ and finite $\lambda$. For simplicity we will work on a flat background $\tilde{R} = 0$. We will begin by working to first order in the deformation. As discussed in the main text, the stress tensor that appears in our deformation excludes the cosmological constant term. We can trade this stress tensor for the total stress tensor $\tilde{T}_{\m\n}^{tot}$ through
\be
\tilde{T}_{\m\n} = \tilde{T}_{\m\n}^{tot} - \m \tilde{g}_{\m\n} e^{2\mathfrak{b}\varphi}\,.
\ee
This means we can rewrite our $\tilde{T}\tilde{\overline{T}}$ operator as 
\be\label{twoterms}
\tilde{T}\tilde{\overline{T}} = (\tilde{T}\tilde{\overline{T}})^{tot} -\f 1 4 \m^2 e^{2\mathfrak{b}\varphi} e^{2\mathfrak{b}\varphi}
\ee
where we regularized the composite operator $(\tilde{T}^\m_\m)^{tot} e^{2b\varphi}$ to vanish, as $\langle \dots(\tilde{T}^\m_\m)^{tot}\dots \rangle = 0$ except for delta-function singularities at coincident points. The combination $(\tilde{T}\tilde{\overline{T}})^{tot}$ has dimension four, while $e^{4\mathfrak{b}\varphi}$ has dimension $4+4\mathfrak{b}^2$. Thus they need different dressings. Using the marginality condition \eqref{tLmarginal} we can dress the stress tensor piece as  

\be
(\tilde T\tilde{\overline{T}})^{tot} e^{-2\xi \varphi}\,,\qquad \xi = \f{q}{2}\left(1 - \sqrt{1-4/q^2}\right)\,
\ee
and the vertex operator piece as 
\be
(e^{2\mathfrak{b}\varphi} e^{2\mathfrak{b}\varphi})e^{-2\mathfrak{b}\varphi} = e^{2\mathfrak{b}\varphi}\,,\qquad \mathfrak{b} = \f q 2 \left(-1+\sqrt{1+4/q^2}\right)
\ee
In Liouville theory we can compute dimensions of fused vertex operators by multiplication, $e^{2\a_1\varphi}e^{2\a_2 \varphi} = e^{2(\a_1+\a_2)\varphi}$. This explains the dressing of the vertex operator piece. The two distinct dressings agree in the semiclassical limit $q\rightarrow \infty$, where we have $\mathfrak{b} \sim \xi \sim 1/q$. 

The first order deformation is therefore given by adding
\be
\d S = \l (\tilde{T}\tilde{\overline{T}})^{tot}e^{-2\xi\varphi} -\f 1 4 \m^2 \l e^{2\mathfrak{b}\varphi}
\ee
to the holographic CFT coupled to timelike Liouville theory \eqref{tLaction}. 

Going beyond first order in these types of flows is generally complicated. The issue is maintaining marginality along the flow, especially after the matter and Liouville sectors mix. We will not study this problem carefully here except to make a few comments. The flow defined here may be more tractable to analyze beyond first order due to the special operator $(\tilde{T}\tilde{\overline{T}})^{tot}$ being dressed. In particular we can define the composite operator as 
\be
(\tilde{T}\tilde{\overline{T}})^{tot}e^{-2\xi\varphi} \defeq L_{-2} \bar{L}_{-2} e^{-2\xi\varphi}\,.
\ee
This definition means the operator we deform by is a Virasoro descendant. Its scaling dimension is given by
\be\label{opdim}
[(\tilde{T}\tilde{\overline{T}})^{tot}e^{-2\xi\varphi}] = [L_{-2} \bar{L}_{-2} e^{-2\xi\varphi}] = [e^{-2\xi\varphi}] + 4.
\ee
Thus the issue of anomalous dimensions is restricted to anomalous dimensions for the vertex operators themselves. These are generically nonzero. At first order they can be calculated by standard methods. We consider the vertex operator $e^{2\a\varphi}$ and the state on $S^1$ that it corresponds to $|e^{2\a\varphi}\rangle$. The change in dimension of the vertex operator is the same as the change in energy of the corresponding state. We can compute the change in the energy by $\delta H \sim -L\langle e^{2\alpha\varphi} | \lambda (\tilde{T}\tilde{\overline{T}})^{tot} e^{-2\xi\varphi}|e^{2\alpha\varphi}\rangle$, where $L$ is the length of $S^1$. This can be mapped to a three-point function on the plane by the cylinder-to-plane conformal map. The correlator between two primary operators and a descendant can be rewritten as a differential operator acting on a correlation function of three primary operators, which is evaluated in the undeformed timelike Liouville theory through the timelike DOZZ formula.
There is some interesting controversy around the timelike DOZZ formula, namely in matching the bootstrap-inspired formula with path integral techniques \cite{Zamolodchikov:2005fy, Kostov:2005kk, Harlow:2011ny, Anninos:2021ene, Chatterjee:2025yzo, Chatterjee:2026zmb} -- we will steer clear of it by not explicitly evaluating the expression above, although we expect a generically nonzero answer and therefore a shift in the operator dimension of $e^{2\alpha\varphi}$.

The fact that the complication of anomalous dimensions is restricted to  the vertex operators suggests that we can maintain conformal symmetry simply by upgrading the dressing of the operators at each stage along the flow. This proposal also rests on the particular nature of the $T\overline{T}$ flow, which picks a single trajectory through coupling space (and so we do not need to consider the generation of additional operators and their dressings).

It may surprise the reader that we have a marginal operator that is a descendant. This can happen because our theory is non-unitary. In particular \eqref{opdim} shows that we need the vertex operator to have dimension $-2$. A marginal, deforming operator that is descendant does not preserve full Virasoro symmetry. Our deforming operator only preserves global $SL(2,C)$ invariance. (The needed upgrade of the stress tensor along the flow is $O(\l)$ and so cannot repair an $O(1)$ violation of Virasoro symmetry.) The precise definition of everything in the quantum theory may lead to an upgrade to a full Virasoro symmetry. Alternatively, we expect introducing the $bc$ ghosts will also lead to a preservation of Virasoro symmetry. This question clearly deserves further study. 

\section{Exact marginality of deformation}\label{app:classmarg}
In this appendix we will discuss the exact marginality of our deformation. First we should discuss what we mean by ``exact." There are two dimensionless expansion parameters in our theory, $c_m$ and $\m \l$. The parameter $c_m$ controls the usual semiclassical limit in pure Liouville theory, and $\m\l$ controls the deformation. The claimed exact marginality of the deformation means that it is quantum-mechanically marginal at finite $c_m$ and $\m\l$, i.e. with respect to both expansion parameters. This is conjectural and assumes that  the coupling of our massive theory to Liouville theory makes sense. In particular, coupling massive theories to the Weyl mode of the metric through Liouville theory is expected to generate conformal field theories, see e.g. Section 4.3 of \cite{Seiberg:1990eb} or Section 2.1 of \cite{Ginsparg:1993is}. As discussed in the previous appendix, however, it is difficult to write down a precise prescription for the quantum theory. 

In the semiclassical limit $c_m \rightarrow \infty$ we can show exact marginality. This basically follows from the fact that $c_m \rightarrow \infty$ in this theory is precisely like an $\hbar \rightarrow 0$ limit, so exact marginality reduces to classical marginality; more on this below. In this limit our deformation can be written as 
\be\label{defA}
\d S = \int d\lambda \int d^2 x \sqrt{\tilde{g}} \,\tilde{T} \tilde{\overline{T}}  e^{-2\Phi}\,.
\ee
As discussed in Appendix \ref{app:qtm}, the two terms in \eqref{twoterms} have the same scaling dimension (equal to four) in the semiclassical limit, so we can pick a single dressing for them. The Ricci scalar curvature term in the deformation \eqref{semideform} simply adds a $\chi \log \l$ term and does not change anything.  There are many ways to see the marginality of \eqref{defA} explicitly. One way is to recall that our deformation descends from the $T\overline{T}$ deformation on the physical metric:
\be\label{trivial}
\sqrt{\tilde{g}}\, \tilde{T} \tilde{\overline{T}}  e^{-2\Phi}  =\sqrt{g} \,T\overline{T} \,.
\ee
The Weyl transformation is defined as 
\be\label{doubleweyl}
\tilde{g}_{\m\n} \rightarrow e^{2\sigma}\tilde{g}_{\m\n}\,,\qquad \Phi \rightarrow \Phi - \sigma
\ee
for general spacetime-dependent $\sigma$. This leaves the physical metric $g_{\m\n} = e^{2\Phi} \tilde{g}_{\m\n}$ invariant, so it will leave any functionals of the physical metric invariant. In particular it leaves $\sqrt{g} T\overline{T}$ invariant, and due to \eqref{trivial} leaves the nonlinear deformation \eqref{defA} invariant. While this is a classical argument, the symmetry \eqref{doubleweyl} is a gauge symmetry and cannot be anomalous, so the symmetry must persist into the quantum theory.\footnote{The reason this argument does not work outside of the semiclassical limit is that we have $g_{\m\n} = e^{2\mathfrak{b}\varphi} \tilde{g}_{\m\n}$ and so $\sqrt{g}\, T\overline{T} = \sqrt{\tilde{g}}\, \tilde{T}\tilde{\overline{T}} e^{-2\mathfrak{b}\varphi}$, which does not provide the correct marginal dressing of our operator. This is because outside of the semiclassical limit the dimensions are not given purely by classical scaling. Nevertheless, there remains an expectation that Weyl invariance is preserved even at finite $c_m$.}

Another way to diagnose the marginality of $\tilde{T}\tilde{\overline{T}}e^{-2\Phi}$ is to compute its scaling dimension. We have $[\tilde{T}\tilde{\overline{T}}e^{-2\Phi}] = 4 + [e^{-2\Phi}]$ as shown in Appendix \ref{app:qtm}. So let's calculate the dimension of a general vertex operator $e^{2\a\Phi}$. Given the semiclassical limit we can do this by saddle point. Since we have a sphere saddle, we can evaluate 
\be
\langle e^{2\a\Phi}\rangle = e^{2\a\Phi_\star} \sim r^{-2\a}\,,
\ee
where $r$ is the radius of the sphere given by the fiducial metric. (In Section \ref{spherebdry} we set $r=1$ but can restore it by $\Phi_\star \rightarrow \Phi_\star + \log r$ in the left hand side of \eqref{phistarsphere}.)  This same answer for the one-point function can be obtained by a different method, for example we can calculate the one-point function for $\alpha = 1$ by differentiating $\log Z$ with respect to $\m$ to bring down $\int d^2 x \sqrt{\tilde{g}} \m e^{2\Phi}$ from the action: $\langle e^{2\Phi}\rangle = \f{1}{4\pi}\partial_\m \log Z$, where we divide by $4\pi$ to eliminate the area of the unit sphere. This gives the same answer for $\a = 1$. (Technically the sphere partition function is divergent in our theory, as is the one-point function, due to the vol$(PSL(2,C))$ factor that remains without introduction of the $bc$ ghosts. So this calculation is technically infinite, but it gives the right scaling property of the vertex operator.) We therefore see that $\Delta[e^{-2\Phi}] = -2$, which gives the classical scaling dimension. Thus the operator $\tilde{T}\tilde{\overline{T}}e^{-2\Phi}$ has dimension $\D = 2$ and is marginal at $c_m \rightarrow \infty$, for arbitrary $\m \l$. 

Here we should make some comments about large-$N$ limits and anomalous dimensions. For general large-$N$ theories, even though the large-$N$ limit is a semiclassical limit, anomalous dimensions of marginal operators are not necessarily suppressed by $N$. One example is the classically marginal $(\phi_i\phi^i)^2$ interaction in 4d, which is marginally irrelevant even at infinite $N$. An actual classical limit $\hbar \rightarrow 0$, on the other hand, should suppress all anomalous dimensions. The distinction is that the traditional $\hbar\rightarrow 0$ limit has $1/\hbar$ sitting in front of the classical action. From the perspective of the path integral this suppresses all quantum corrections.  One can try to employ the same line of logic for the large-$N$ limit, but most large-$N$ theories do not have a factor of $N$ sitting in front of the action. Instead, there are $N$ fields that appear in the action which are summed over.\footnote{To get an explicit $N$ in front, one needs to pass to a ``collective" or ``master" field description. The collective fields in holographic CFTs are the bulk variables, and they have $N \sim 1/G$ in front of the action. In vector models the collective field is given by the bilocal field $\psi(x,y) \sim \phi_i(x) \phi^i(y)$ \cite{Cornwall:1974vz}, see also \cite{Jevicki:1980zg}. In terms of this variable one gets an effective action with an overall $N$ in front, and the semiclassical limit can be analyzed by saddle points as an ordinary $\hbar\rightarrow 0$ limit. As a result, the collective fields have scaling dimension given by classical scaling.} So large $N$ does not necessarily suppress quantum effects. In Liouville theory the large parameter, $c_m \sim 1/G$, sits in front of the action. This is true also of the $\tilde{T}\tilde{\overline{T}}$ term for semiclassical states with $T_{\m\n} \sim 1/G$. So  the semiclassical limit $c_m \rightarrow \infty$ is like a standard $\hbar\rightarrow 0$ limit that suppresses all loop corrections. This explains why the classical scaling dimensions of our vertex operators are exact in $\m\l$ at infinite $c_m$. 

Given that in our deformed Liouville theory $c_m$ appears precisely as $1/\hbar$, we expect classical scaling to hold in the $c_m \rightarrow \infty$. This is another way to see why we obtained the classical dimensions for the vertex operators, even though we have a deformation that can be treated quantum-mechanically. 

We can also see the Weyl invariance \eqref{doubleweyl} at the level of the classical deformation by direct calculation. This invariance is guaranteed by \eqref{trivial}, but it might be worth seeing the mechanics of the invariance in the frame of the fiducial metric. Doing calculations with an obvious answer in a non-obvious gauge is not an amazing idea, but let's do it anyway. The argument is inductive, so we will begin at first order in $\l$, which is the only order in which we will be very explicit. At first order, note that \eqref{trivial} implies that the zeroeth order stress tensor $\tilde{T}_{\mu\nu}$ includes both the matter stress tensor on the fiducial metric and the Wess-Zumino stress tensor,
\be\label{bothstress}
\tilde{T}_{\mu\nu}[\tilde{g},\Phi]=\tilde{T}^m_{\mu\nu}[\tilde{g}]+\tilde T^{WZ}_{\mu\nu}[\tilde{g},\Phi].
\ee
 The Wess-Zumino action is 
 \be
\label{WZaction}
S_{WZ}[\tilde{g}, \Phi]=-\f{1}{4\pi\mathfrak{b}^2}\int d^2x\sqrt{\tilde{g}}\left[(\tilde{\nabla}\Phi)^2+\tilde{R}\Phi\right]
 \ee
and has stress tensor
\be
\tilde T_{\mu\nu}^{WZ}[\tilde{g}, \Phi]=-\f{2}{\sqrt{\tilde{g}}}\f{\delta S_{WZ}}{\delta\tilde{g}^{\mu\nu}}=\f{1}{2\pi\mathfrak{b}^2}\left(\tilde{\nabla}_\mu\Phi\tilde{\nabla}_\nu\Phi-\tilde{\nabla}_\nu\tilde{\nabla}_\mu\Phi+\tilde{g}_{\mu\nu}\tilde{\square}\Phi-\f{1}{2}\tilde{g}_{\mu\nu}(\tilde{\nabla}\Phi)^2\right),\label{twz}
\ee
where $\mathfrak{b}^2\approx 6/c_m$. 

We will first check the transformation of \eqref{bothstress} by checking the transformation of each piece. We have
\be
\tilde{T}_{\m\n}^m[\tilde{g}] = -\f{2}{\sqrt{\tilde{g}}} \f{\delta S_m[\tilde{g}]}{\delta \tilde{g}^{\m\n}}\,.
\ee
The denominator $\sqrt{\tilde{g}}\, \delta\tilde{g}^{\m\n}$ is invariant under the Weyl transformation, whereas $S_m$ transforms via the Wess-Zumino action:
\be\label{tm}
\tilde{T}_{\m\n}^m[\tilde{g}] = -\f{2}{\sqrt{\tilde{g}}} \f{\delta S_m[\tilde{g}]}{\delta \tilde{g}^{\m\n}} \longrightarrow  -\f{2}{\sqrt{\tilde{g}}} \f{\delta S_m[e^{2\sigma}\tilde{g}]}{\delta \tilde{g}^{\m\n}} = -\f{2}{\sqrt{\tilde{g}}} \f{\delta (S_m[\tilde{g}]+S_{WZ}[\tilde{g},\sigma]}{\delta \tilde{g}^{\m\n}} = \tilde{T}_{\m\n}^m[\tilde{g}] + \tilde{T}_{\m\n}^{WZ}[\tilde{g},\sigma] 
\ee
Now we check the transformation of $T_{\m\n}^{WZ}[\tilde{g},\Phi]$. Under $\tilde{g}_{\mu\nu}\rightarrow e^{2\sigma}\tilde{g}_{\mu\nu}$ we have
	\begin{equation}
		\begin{aligned}
				&\tilde{T}^{WZ}_{\mu\nu}[e^{2\sigma}\tilde{g},\Phi]\\
				&=\frac{1}{2\pi \mathfrak{b}^2}\left(\tilde{\nabla}_\mu\Phi\tilde{\nabla}_\nu\Phi-\tilde{\nabla}_\nu\tilde{\nabla}_\mu\Phi+\tilde{g}_{\mu\nu}\tilde{\square}\Phi-\frac{1}{2}\tilde{g}_{\mu\nu}(\tilde{\nabla}\Phi)^2+\tilde{\nabla}_\mu\Phi\tilde{\nabla}_\nu\sigma+\tilde{\nabla}_\mu\sigma\tilde{\nabla}_\nu\Phi-\tilde{g}_{\mu\nu}\tilde{\nabla}_\rho\Phi\tilde{\nabla}^\rho\sigma\right).
		\end{aligned}
	\end{equation}
 Doing a further $\Phi\rightarrow\Phi-\sigma$ we get
	\begin{equation}\label{twztransform}
	\begin{aligned}
		&\tilde{T}^{WZ}_{\mu\nu}[e^{2\sigma}\tilde{g},\Phi-\sigma]\\
		&=\frac{1}{2\pi \mathfrak{b}^2}\left(\tilde{\nabla}_\mu\Phi\tilde{\nabla}_\nu\Phi-\tilde{\nabla}_\nu\tilde{\nabla}_\mu\Phi+\tilde{g}_{\mu\nu}\tilde{\square}\Phi-\frac{1}{2}\tilde{g}_{\mu\nu}(\tilde{\nabla}\Phi)^2-\tilde{\nabla}_\mu\sigma\tilde{\nabla}_\nu\sigma+\tilde{\nabla}_\nu\tilde{\nabla}_\mu\sigma-\tilde{g}_{\mu\nu}\tilde{\square}\sigma+\frac{1}{2}\tilde{g}_{\mu\nu}(\tilde{\nabla}\sigma)^2\right).
	\end{aligned}
\end{equation}
Thus we get 
\be
\tilde{T}_{\m\n}^{WZ}[\tilde{g},\Phi] \longrightarrow \tilde{T}^{WZ}_{\mu\nu}[e^{2\sigma}\tilde{g},\Phi-\sigma] =  \tilde{T}_{\m\n}^{WZ}[\tilde{g},\Phi] - \tilde{T}_{\m\n}^{WZ}[\tilde{g},\sigma]\,.
\ee
Combining this with \eqref{tm}, we see that
\begin{equation}
\tilde{T}_{\mu\nu}^m[e^{2\sigma}\tilde{g}]+\tilde{T}_{\mu\nu}^{WZ}[e^{2\sigma}\tilde{g},\Phi-\sigma] = \tilde{T}^m_{\mu\nu}[\tilde{g}]+\tilde{T}_{\mu\nu}^{WZ}[\tilde{g},\Phi]\,,
\end{equation}
i.e. our stress tensor $\tilde{T}_{\m\n}$ is invariant under \eqref{doubleweyl}.
Now we can consider the operator 
\be\label{fullop}
\sqrt{\tilde{g}} \,\tilde{T}\tilde{\overline{T}} e^{-2\Phi} =\f 1 8 \sqrt{\tilde{g}}\, \tilde{T}_{\m\n} \tilde{T}_{\a\b}(\tilde{g}^{\m \a}\tilde{g}^{\n\b} - \tilde{g}^{\m\n} \tilde{g}^{\a\b})e^{-2\Phi}.
\ee
The piece $\tilde{T}_{\m\n} \tilde{T}_{\a\b}$ is invariant as just shown. The transformation of the rest is
\be\label{gtrans}
\sqrt{\tilde{g}} (\tilde{g}^{\m \a}\tilde{g}^{\n\b} - \tilde{g}^{\m\n} \tilde{g}^{\a\b}) e^{-2\Phi}\longrightarrow \sqrt{e^{4\sigma}\tilde{g}} (e^{-4\sigma}\tilde{g}^{\m \a}\tilde{g}^{\n\b} - e^{-4\sigma}\tilde{g}^{\m\n} \tilde{g}^{\a\b})e^{-2\Phi+2\sigma}\,.
\ee
So this piece is also invariant, and we therefore see that our deformation \eqref{defA} is invariant under Weyl transformations \eqref{doubleweyl}. This means our action at first order in $\l$ is Weyl invariant, since all other pieces of the action are trivially Weyl invariant. 

Now we assume the action at $O(\l^n)$ is Weyl invariant. The stress tensor at this order can be very messy, so we will not do an explicit calculation. Instead, we will argue from general principles. The cosmological constant term is trivially Weyl invariant, so the action that excludes this term is also Weyl invariant. The stress tensor $\tilde{T}_{\m\n}$ is calculated by varying the action excluding this cosmological constant term with respect to the fiducial metric, see below \eqref{paramids}. Since the action is Weyl invariant, and $\sqrt{\tilde{g}} \, \delta \tilde{g}^{\m\n}$ is Weyl invariant, the stress tensor at this order is Weyl invariant. In fact this Weyl invariance follows directly from our previous general argument, since $\tilde{T}_{\m\n}[\tilde{g},\Phi] = T_{\m\n}[g]$, i.e. the fiducial stress tensor can be thought of as a functional just of the physical metric. In any case, by the same argument in \eqref{fullop}-\eqref{gtrans}, the full deformation at this order is Weyl invariant, making the action at $O(\l^{n+1})$ Weyl invariant. This completes the induction.

\section{Deformed Liouville equation of motion}\label{app:ttbarwz}
In this appendix we want to argue that the equation of motion for $\Phi$ in the semiclassical limit defined by \eqref{timelikesemi}-\eqref{semideform} is given by
\be\label{lioueom}
2\tilde{\square}\Phi-\tilde{R}+8\pi\mathfrak{b}^2\mu e^{2\Phi}-8\pi\mathfrak{b}^2\lambda \tilde{T}\tilde{\overline{T}}(\lambda) e^{-2\Phi}=0.
\ee
The first three terms follow immediately by differentiation. The fourth term needs some explanation. To understand it, let's start with the traditional $T\overline{T}$ deformation, which is defined on the physical metric as
\be
\f{\partial S}{\partial \lambda}=\int d^2x \sqrt{g}\left(T\overline{T}+\frac{c_m}{48\pi\lambda}R\right).
\ee
Defining the effective action $W=-\log Z$, this leads to
\be
\f{\partial W}{\partial \lambda}=\int d^2x \sqrt{g}\left(\langle T\overline{T}\rangle+\frac{c_m}{48\pi\lambda}R\right).
\ee
This derivative of the effective action with respect to the only scale, $\lambda$, generates scale transformations  as (see e.g. section 3 of \cite{Hartman:2018tkw})
\be
-2\lambda \f{\partial W}{\partial \lambda}=\int d^2x \sqrt{g}\langle T^{\mu}_\mu\rangle\,.
\ee
The variation of the effective action with respect to the Weyl mode also generates scale transformations:
\be
\f{\delta W}{\delta \Phi}= \sqrt{g}\langle T^{\mu}_\mu\rangle\,.
\ee
Thus we have
\be
\int d^2x \f{\delta W}{\delta \Phi} = -2\lambda \f{\partial W}{\partial \lambda}\,.
\ee
This can be translated to a local statement by upgrading the deformation parameter $\lambda\rightarrow\lambda(x)$, and writing $\f{\partial W}{\partial \lambda}\rightarrow\frac{\delta W}{\delta \lambda}$ to get
\be
\frac{\delta W}{\delta \Phi}=-2\lambda(x) \frac{\delta W}{\delta \lambda}.\label{dlambdadphi}
\ee
Next we write the action for the standard $T\overline{T}$-deformed theory in terms of the fiducial metric:
\be
\begin{aligned}\label{bndryaction}
    S& =S_m[\tilde{g}]+S_{WZ}[\tilde{g}]+\int \mathcal{D}\lambda \int d^2x\sqrt{\tilde{g}}\,\tilde{T}\tilde{\overline{T}}[\lambda(x)]e^{-2\Phi}+\f{c_m}{48\pi}\int d^2x\sqrt{\tilde{g}}\:\log\lambda(x) \left(\tilde{R}-2\tilde{\square}\Phi\right)
    .
\end{aligned}
\ee
$S_{WZ}$ is the Wess-Zumino action encapsulating the anomaly of the seed CFT and is given in \eqref{WZaction}.
Taking the $\Phi$ variation of \eqref{bndryaction} gives 
\be\label{dphi}
\f{\delta S}{\delta \Phi}=-\f{c_m}{24\pi}\sqrt{g}R+\frac{\delta}{\delta\Phi}\left(\int \mathcal{D}\lambda \int d^2x\sqrt{\tilde{g}}\,\tilde{T}\tilde{\overline{T}}[\lambda(x)]e^{-2\Phi}\right)-\f{c_m}{24\pi}\sqrt{\tilde{g}}\:\tilde{\square}\log\lambda(x),
\ee
where the first term in the right hand side comes from the fact that the $\Phi$ variation of the Wess-Zumino action gives the anomaly.
Taking the $\lambda$ derivative of \eqref{bndryaction} leads to 
\be\label{dlambda}
\begin{aligned}
   \f{\delta S}{\delta\lambda}=&\sqrt{\tilde{g}}\,\tilde{T}\tilde{\overline{T}}[\lambda(x)]e^{-2\Phi}+\f{c_m}{48\pi\lambda(x)}\sqrt{\tilde{g}} \left(\tilde{R}-2\tilde{\square}\Phi\right) 
\end{aligned}
\ee
Plugging \eqref{dphi} and \eqref{dlambda} into \eqref{dlambdadphi} and using $\sqrt{g}R=\sqrt{\tilde{g}}(\tilde{R}-2\tilde{\square}\Phi)$, we see that 
\be\label{phivar}
\begin{aligned}
\frac{\delta}{\delta\Phi}\left(\int \mathcal{D}\lambda \right.&\left.\int d^2x\sqrt{\tilde{g}}\,\tilde{T}\tilde{\overline{T}}[\lambda(x)]e^{-2\Phi}\right) \\
&=-2\lambda \sqrt{\tilde{g}}\,\tilde{T}\tilde{\overline{T}}[\lambda(x)]e^{-2\Phi}+\f{c_m}{24\pi}\sqrt{\tilde{g}}\:\tilde{\square}\log\lambda(x).
\end{aligned}
\ee
Now if we turn off the coordinate dependence of the deformation parameter $\lambda(x)\rightarrow\lambda$, \eqref{phivar} implies that the contribution of the $\tilde{T}\tilde{\overline{T}}e^{-2\Phi}$ term to the $\Phi$ equation of motion at all orders in $\lambda$ is given by 
\be\label{fourthterm}
-8\pi\mathfrak{b}^2\lambda \tilde{T}\tilde{\overline{T}}(\lambda)e^{-2\Phi},
\ee
as claimed in \eqref{lioueom}.

This argument was a bit roundabout, so let's do an explicit calculation, at least at first order in $\l$. We write the deformation \eqref{semideform} as
\be
S_{def}=\f{c_m\chi}{12}\log \lambda+\lambda\int d^2x\sqrt{\tilde{g}}\,\tilde{T}\tilde{\overline{T}}e^{-2\Phi}.
\ee
Varying this with respect to $\Phi$ gives
\be
\delta S_{def}=-2\lambda\int d^2 x\sqrt{\tilde{g}}\,\tilde{T}\tilde{\overline{T}}e^{-2\Phi}\delta\Phi+\lambda \int d^2x\sqrt{\tilde{g}}\:\delta(\tilde{T}\tilde{\overline{T}})e^{-2\Phi}.\label{sdefvar}
\ee
The $\tilde{T}\tilde{\overline{T}}$ operator is defined by 
\be
\tilde{T}\tilde{\overline{T}}=\f{1}{8}\left(\tilde{T}_{\mu\nu}\tilde{T}^{\mu\nu}-(\tilde{T}_\mu^\mu)^2\right),
\ee
where $\tilde{T}_{\mu\nu}$ is the stress tensor of the matter sector plus the Wess-Zumino sector, 
\be
\tilde{T}_{\mu\nu}=\tilde{T}^m_{\mu\nu}+\tilde T^{WZ}_{\mu\nu}.
\ee
The Wess-Zumino stress tensor comes from varying \eqref{WZaction} and is given in \eqref{twz}. The $\tilde T\tilde{\overline{T}}$ operator is then given by
\be
\begin{aligned}
    \tilde{T}\tilde{\overline{T}}=&\,\f{1}{8}\Bigg(\tilde{T}^m_{\mu\nu}\tilde{T}^{m\:\mu\nu}-(\tilde{T}_\mu^{m\:\mu})^2+\f{\tilde{T}^m_{\mu\nu}\tilde{\nabla}^\mu\Phi\tilde{\nabla}^\nu\Phi}{\pi\mathfrak{b}^2}-\f{\tilde{T}^m_{\mu\nu}\tilde{\nabla}^\nu\tilde{\nabla}^\mu\Phi}{\pi \mathfrak{b}^2}-\f{\tilde{T}_\mu^{m\:\mu}(\tilde{\nabla}\Phi)^2}{2\pi\mathfrak{b}^2}\\
    &+\f{(\tilde{\nabla}\Phi)^4}{8\pi^2\mathfrak{b}^4}+\f{\tilde{\square}\Phi(\tilde{\nabla}\Phi)^2}{4\pi^2\mathfrak{b}^4}-\f{\tilde{\nabla}_\mu\tilde{\nabla}_\nu\Phi\tilde{\nabla}^\mu\Phi\tilde{\nabla}^\nu\Phi}{2\pi^2\mathfrak{b}^4}-\f{\tilde{\square}\Phi\tilde{\square}\Phi}{4\pi^2\mathfrak{b}^4}+\f{\tilde{\nabla}_\mu\tilde{\nabla}_\nu\Phi\tilde{\nabla}^\mu\tilde{\nabla}^\nu\Phi}{4\pi^2\mathfrak{b}^4}\Bigg).\label{ttmwz}
\end{aligned}
\ee
Using \eqref{ttmwz}, the second term in the variation \eqref{sdefvar} is given by
\be
\begin{aligned}
    \lambda \int &d^2x\sqrt{\tilde{g}}\:\delta(\tilde{T}\tilde{\overline{T}})e^{-2\Phi}=  \f{\lambda}{32\pi^2\mathfrak{b}^4} \int d^2x\sqrt{\tilde{g}}e^{-2\Phi}\Bigg(\tilde{R}\tilde{\Box}\Phi-2(\tilde{\nabla}\Phi)^2\tilde{R}+\tilde{\nabla}_\mu\Phi\tilde{\nabla}^\mu\tilde{R} 
    \\
    & -8\pi\mathfrak{b}^2 \tilde{T}_\mu^{m\:\mu}(\tilde{\nabla}\Phi)^2+4\pi\mathfrak{b}^2\tilde{T}_\mu^{m\:\mu}\tilde{\square}\Phi+8\pi\mathfrak{b}^2\tilde{\nabla}^\mu \tilde{T}^m_{\mu\nu}\tilde{\nabla}^\nu\Phi+4\pi\mathfrak{b}^2\tilde{\nabla}_\mu \tilde{T}^{m\:\nu}_\nu\tilde{\nabla}^\mu\Phi\\
    &-4\pi\mathfrak{b}^2\tilde{\nabla}^\nu\tilde{\nabla}^\mu \tilde{T}_{\mu\nu}^m\Bigg)\delta\Phi.\label{varwz}
\end{aligned}
\ee
Now consider the covariant conservation of the matter stress tensor and the stress tensor of the timelike Liouville theory:
\be
\tilde{\nabla}^\mu \tilde{T}^m_{\mu\nu}=0,\quad \tilde{\nabla}^\mu \tilde{T}^{tL}_{\mu\nu}=\tilde{\nabla}^{\mu}\tilde{T}^{WZ}_{\mu\nu}+\mu\tilde{\nabla}^\mu(\tilde{g}_{\mu\nu}e^{2\Phi})=0.
\ee
The second term in the stress tensor of the timelike Liouville theory comes from the cosmological constant term in the action. Putting these conservations together we get
\be\label{fullcons}
\tilde{\nabla}^\mu\left( \tilde{T}^m_{\mu\nu}+\tilde{T}^{WZ}_{\mu\nu}+\mu\tilde{g}_{\mu\nu}e^{2\Phi}\right)=0.
\ee
Next we remember that the full stress tensor for the boundary theory is traceless:
\be
\tilde{T}^{full\:\mu}_{\mu}=\tilde{T}^{m\:\mu}_\mu+\tilde{T}^{WZ\:\mu}_\mu+2\mu e^{2\Phi}=0.
\ee
So we can add a zero term proportional to this trace to \eqref{fullcons} and write
\be
\begin{aligned}
 \tilde{\nabla}^\mu\left( \tilde{T}^m_{\mu\nu}+\tilde{T}^{WZ}_{\mu\nu}+\mu\tilde{g}_{\mu\nu}e^{2\Phi}\right)&-\tilde{\nabla}_\nu\Phi\left(\tilde{T}^{m\:\mu}_\mu+\tilde{T}^{WZ\:\mu}_\mu+2\mu e^{2\Phi}\right)=0\\
 \implies \tilde{\nabla}^\mu\left( \tilde{T}^m_{\mu\nu}+\tilde{T}^{WZ}_{\mu\nu}\right)&-\tilde{\nabla}_\nu\Phi\left(\tilde{T}^{m\:\mu}_\mu+\tilde{T}^{WZ\:\mu}_\mu\right)=0.
\end{aligned}
\ee
We can multiply this expression by $e^{-2\Phi}$ and take a $\tilde{\nabla}^\nu$ derivative of it to get\footnote{The manipulations leading to \eqref{cons} might seem arbitrary, however, it is worth noting that they are equivalent to the covariant conservation of the full stress tensor on the physical metric:
\be
\nabla^\mu T^{full}_{\mu\nu}=e^{-2\Phi}\left(\tilde{\nabla}^\mu \tilde{T}^m_{\mu\nu}+\tilde{\nabla}^\mu \tilde T^{WZ}_{\mu\nu}-\tilde{\nabla}_\nu\Phi \tilde{T}^{m\:\mu}_\mu-\tilde{\nabla}_\nu \Phi \tilde T^{WZ\:\mu}_\mu\right)=0.
\ee}:
\be
\begin{aligned}
   0&=\tilde{\nabla}^\nu\left(e^{-2\Phi}\left(\tilde{\nabla}^\mu \tilde{T}^m_{\mu\nu}+\tilde{\nabla}^\mu \tilde T^{WZ}_{\mu\nu}-\tilde{\nabla}_\nu\Phi \tilde{T}^{m\:\mu}_\mu-\tilde{\nabla}_\nu \Phi \tilde T^{WZ\:\mu}_\mu\right)\right)\\
&=-\f{e^{-2\Phi}}{4\pi \mathfrak{b}^2}\Bigg(\tilde{R}\tilde{\Box}\Phi-2(\tilde{\nabla}\Phi)^2\tilde{R}+\tilde{\nabla}_\mu\Phi\tilde{\nabla}^\mu\tilde{R} 
     -8\pi\mathfrak{b}^2 \tilde{T}_\mu^{m\:\mu}(\tilde{\nabla}\Phi)^2+4\pi\mathfrak{b}^2\tilde{T}_\mu^{m\:\mu}\tilde{\square}\Phi\\&+8\pi\mathfrak{b}^2\tilde{\nabla}^\mu \tilde{T}^m_{\mu\nu}\tilde{\nabla}^\nu\Phi+4\pi\mathfrak{b}^2\tilde{\nabla}_\mu \tilde{T}^{m\:\nu}_\nu\tilde{\nabla}^\mu\Phi-4\pi\mathfrak{b}^2\tilde{\nabla}^\nu\tilde{\nabla}^\mu \tilde{T}_{\mu\nu}^m\Bigg).\label{cons}
\end{aligned}
\ee
In the second equality we used \eqref{twz}. \eqref{cons} is proportional to the term in the parenthesis in \eqref{varwz}, so we can deduce 
\be
\lambda \int d^2x\sqrt{\tilde{g}}\delta(\tilde{T}\tilde{\overline{T}}) e^{-2\Phi}=0\,.
\ee
As a result, \eqref{sdefvar} reduces to 
\be
\delta S_{def}=-2\lambda\int d^2 x\sqrt{\tilde{g}}\tilde{T}\tilde{\overline{T}}e^{-2\Phi}\delta\Phi\,.
\ee
Thus we can explicitly see, at least at first order in $\lambda$, that the Wess-Zumino sector does not contribute to the variation of $\tilde{T}\tilde{\overline{T}}$, and we recover the contribution to the equation of motion \eqref{fourthterm}.

\section{$T\overline{T}$-deformed CFT$_2$ on a torus}\label{app:ttbartorus}
In this appendix we will derive the nonperturbatively flowed matter $T\overline{T}$ operator \eqref{ttbartorus} used in the calculation of the torus partition function. The first step will be to write down the flowed partition function, from which the stress tensor can be calculated. We will obtain the partition function in two ways: through an integral transform \cite{Dubovsky:2018bmo} and through the requirement of modular covariance \cite{Datta:2018thy}. 

We consider the $T\overline{T}$ deformation of a CFT$_2$ with central charge $c$ defined on a torus of spatial circle size $2\pi R$. The partition function for the CFT$_2$ is defined by 
\begin{equation}
    Z_{CFT}(\t_1,\t_2)=\sum_n \exp\left(2\pi i\left(\tau_1 J_n+i\tau_2 M_n\right)\right),\label{cftz}
\end{equation}
where $\tau = \tau_1 + i\tau_2$ is the modular parameter of the torus, $M_n=\Delta_n+\overline{\Delta}_n-c/12$, and $J_n=\Delta_n-\overline{\Delta}_n$. The modular parameters are related to the temperature and angular velocity of the CFT$_2$ by $\beta'=2\pi R\t_2$ and $i\beta'\Omega'=2\pi \t_1$. Using the modular invariance of the partition function, we can use the $S:\t\rightarrow-1/\tau$ transform to see that in the high temperature limit, $\beta'\rightarrow 0$, the partition function is dominated by the vacuum state in the dual channel and hence can be approximated by
\begin{equation}\label{hightempcft}
Z_{CFT}(\t_1,\t_2)\approx \exp\left(\frac{\pi c\t_2}{6(\t_1^2+\t_2^2)}\right).
\end{equation} 
This can be extended to intermediate temperatures as discussed in the main text. The $T\overline{T}$-deformed partition function can be found by the integral transform \cite{Dubovsky:2018bmo} (we will use the conventions of \cite{Callebaut:2019omt} up to a sign flip $\lambda\rightarrow -\lambda$)
\begin{equation}\label{inttransf}
    Z_{\text{deformed}}(\lambda,\sigma_1,\sigma_2)=-\f{\sigma_2}{\pi \lambda}\int_{\mathbb{H}}\frac{d\t_1 d\t_2}{\t_2^2}e^{\f{4\pi^2 R^2}{\lambda \t_2}|\t-\sigma|^2}Z_{CFT}(\t_1,\t_2),
\end{equation}
where $\s = \sigma_1+ i\sigma_2$ is the modular parameter for the deformed theory. Convergence of this integral requires $\l < 0$. To see that this integral transform is correct, we can use \eqref{cftz} and evaluate the integral transform term by term to obtain
\begin{equation}
    Z_{\text{deformed}}(\lambda,\sigma_1,\sigma_2)=\sum_n \exp\left(2\pi i \sigma_1 J_n -\sigma_2 \frac{8\pi^2R^2}{\lambda}\left(1-\sqrt{1-\frac{\lambda M_n}{2\pi R^2}+\f{\lambda^2 J_n^2}{16\pi^2 R^4}}\right)\hspace{-1mm}\right).\label{deformedzexpansion}
\end{equation}
Notice this is just the usual partition sum for a theory with energies and momenta given by those of a $T\overline{T}$-deformed theory, justifying \eqref{inttransf}. 

Plugging \eqref{hightempcft} into \eqref{inttransf}, we can evaluate the integral by saddle point to get
\begin{equation}
    \log Z_{\text{deformed}}(\lambda,\beta,\Omega)\approx -\frac{4\pi R\beta(1-R^2\Omega^2)-2\pi R\sqrt{\f{2}{3}\left(1-R^2\Omega^2\right)(\pi c\lambda+6\beta^2(1-R^2\Omega^2))}}{\lambda(1-R^2\Omega^2)},\label{deformedzapprox}
\end{equation}
where we used $2\pi R\sigma_2=\beta$, and $2\pi \sigma_1=i\beta \Omega$.

We can also get this expression using the modular covariance of the $T\overline{T}$-deformed partition function \cite{Datta:2018thy}: 
\begin{equation}
    Z_{\text{deformed}}\left(\f{\lambda}{R^2},\sigma\right) = Z_{\text{deformed}}\left(\f{1}{|c\sigma+d|^2}\f{\lambda}{R^2},\frac{a\sigma+b}{c\sigma+d}\right),
\end{equation}
where we formed the dimensionless coupling $\l/R^2$.  The modular S transform is given by $a=d=0$, $b=-c=1$, under which 
\begin{equation}
    \sigma_1\rightarrow -\f{\sigma_1}{|\sigma|^2},\quad \sigma_2\rightarrow \f{\sigma_2}{|\sigma|^2},\quad \f{\lambda}{R^2}\rightarrow \f{\lambda}{R^2}\f{1}{|\sigma|^2}\,.
\end{equation}
Taking $|\s|\rightarrow 0$ projects us to the vacuum in the dual channel, which is given by  $M_n=-c/12$, $J_n=0$. This means \eqref{deformedzexpansion} is given by \eqref{deformedzapprox} at high temperature. We can now analytically continue $\l$ to positive values, which matches the sign in the main text of the paper.

From this expression we can calculate the needed $T\overline{T}$ operator by using the thermodynamic relations for energy density, momentum density, and pressure, corresponding to $T_{tt}=-(2\pi R)^{-1} (\p_\b -\Omega \b^{-1}\p_\Omega ) \log Z$, $T_{tx} = (2\pi R^2\b)^{-1}\p_\Omega \log Z$, and $T_{xx}= (2\pi \b)^{-1}\p_R \log Z$, respectively. This gives \eqref{ttbartorus}. Alternatively, we can use 
\begin{equation}
    \f{\partial S}{\partial \lambda}=\int d^2x\sqrt{g}\,T\overline{T}
\end{equation}
to calculate
\begin{equation}\label{ttbtorusbndry}
    -\f{1}{2\pi R\beta }\f{\partial}{\partial \lambda}\log Z_{\text{deformed}}=T\overline{T} = \f{2 }{\l^2  }\left(\f{1-R^2\Omega^2+ \pi c \l  /(12\b^2)}{\sqrt{(1-R^2\Omega^2)(1-R^2\Omega^2 + \pi c \l /(6\b^2)})}-1\right).
\end{equation}
To match our conventions in the main text, let's rewrite $T\overline{T}$ on a torus that is parameterized as in \eqref{rotbdrymetric2}: $ds^2 = \Lambda^2(d\tilde{\t}^2 + d\phi^2)$. To do so we should replace $\beta \rightarrow \Lambda \tilde\b$ and $R \rightarrow \L$, which gives $\tilde\Omega \rightarrow \Lambda \Omega$. Plugging into the expression above gives
\be
T\overline{T}=\f{2 }{\l^2   }\left(\f{1-\tilde{\Omega}^2+ \pi c \l  /(12\L^2\tilde{\b}^2)}{\sqrt{(1-\tilde{\Omega}^2)(1-\tilde{\Omega}^2 + \pi c \l/(6\L^2\tilde{\b}^2)})}-1\right).
\ee
Now we use $\Lambda = e^{\Phi}$ and \eqref{physfid} to get
\be\label{ttbartildetorus}
\tilde{T}\tilde{\overline{T}}=e^{4\Phi}T\overline{T}=\f{2 }{\l^2 e^{-4\Phi}  }\left(\f{1-\tilde{\Omega}^2+ \pi c \l e^{-2\Phi}  /(12\tilde{\b}^2)}{\sqrt{(1-\tilde{\Omega}^2)(1-\tilde{\Omega}^2 + \pi c \l e^{-2\Phi}/(6\tilde{\b}^2)})}-1\right).
\ee
As explained below \eqref{semideform}, we can also obtain $\tilde{T}\tilde{\overline{T}}$ by taking $T\overline{T}$ and mapping $\l \rightarrow \l e^{-2\Phi}$ and trading the physical metric for the fiducial metric (i.e. setting $\Lambda = 1$). This gives \eqref{ttbartildetorus}.

\small
\bibliographystyle{jhep}
\bibliography{references.bib}

\end{document}